\documentclass[%
 aip,
 amsmath,amssymb,
 reprint,%
]{revtex4-1}

\usepackage{graphicx}% Include figure files
\usepackage{dcolumn}% Align table columns on decimal point
\usepackage{bm}% bold math
\usepackage{comment}
\usepackage{xcolor}

\usepackage[utf8]{inputenc}
\usepackage[T1]{fontenc}
\usepackage{mathptmx}

\begin{document}

\preprint{AIP/123-QED}

\title[Autoionization Dynamics]{Quantum-classical Dynamics of Vibration-Induced Autoionization\\ in Molecules}% Force line breaks with \\

\author{Kevin Issler}%
\author{Roland Mitri\'{c}}
 \email{roland.mitric@uni-wuerzburg.de}
\author{Jens Petersen}
 \email{jens.petersen@uni-wuerzburg.de}
\affiliation{%
 Institut f\"{u}r physikalische und theoretische Chemie, Julius-Maximilians-Universit\"{a}t W\"{u}rzburg, Emil-Fischer-Str. 42, 97074 W\"{u}rzburg, Germany 
 }%

\date{\today}% It is always \today, today,

\begin{abstract}
We present a novel method for the simulation of the vibration-induced autoionization dynamics in molecular anions in the framework of the quantum-classical surface hopping approach. 
Classical trajectories starting from quantum initial conditions are propagated on a quantum-mechanical potential energy surface while allowing for autoionization through transitions into discretized continuum states. These transitions are induced by the couplings between the electronic states of the bound anionic system and the electron-detached system composed of the neutral molecule and the free electron.
A discretization scheme for the detached system is introduced and a set of formulae is derived which enables the approximate calculation of couplings between the bound and free-electron states.

We demonstrate our method on the example of the anion of vinylidene, a high-energy isomer of acetylene, for which detailed experimental data is available. 
Our results provide information on the time scale of the autoionization process and give an insight into the energetic and angular distribution of the ejected electrons as well as into the associated changes of the molecular geometry. 
We identify the formation of structures with reduced C-C bond lengths and T-like conformations through bending of the CH$_2$ group with respect to the C-C axis and point out the role of autoionization as a driving process for the isomerization to acetylene. 
\end{abstract}

\maketitle

\section{\label{sec:Intro}Introduction}
Molecular anions are often characterized by small electron binding energies, and it is not uncommon that the additional electron(s) are not strictly bound at all, the anionic molecular system thus being in a metastable or "quasi-bound" state that after a finite lifetime decays by ejecting an electron in a process termed autoionization.\cite{reviewsimons,anionbook}
This seemingly quite exotic process is in fact relevant in a variety of areas. Notably, the capture of slow electrons by biological molecules such as nucleobases may populate metastable states that decay by autoionization, accompanied by chemical transformations of the molecule.\cite{uracil,uracilthymine,dnatonzani,dnareview} 
In DNA, electron attachment and subsequent autoionization can occur both at nucleobases and the phosphate-deoxyribose backbone, causing single and, through consecutive reaction of fragments, double strand breaks.\cite{dnareview,dnascience,dnaprl}
Since the electrons can, e.g., be produced by primary ionization due to nuclear radiation, this makes autoionization processes a key step in the radiation damage of biological systems. 
Furthermore, low-energy electron generation via intermolecular coulombic decay, first discovered in noble gas clusters\cite{hergenhahnNe} and liquid water\cite{hergenhahnWater}, occurs in biology at the FADH$^-$ cofactor where it enables the mechanism of photolesion repair in DNA by photolyases.\cite{harbach}
Autoionization is also an important part of the manifold generation and dissipation mechanisms in solvent clusters with excess electrons,\cite{oliverSolvE,solvelectrons} which in turn play a significant role in processes such as nucleation and aerosol formation in the upper atmosphere\cite{ArnoldSolvE} or in the formation of solvated electrons in living organisms upon UV irradiation of riboflavin and its derivatives.\cite{getoffSolvE}
Moreover, the formation of anions that can undergo autoionization is also an important mechanism in the creation of complex molecules in interstellar space.\cite{astro}

The metastable states or resonances from which autoionization takes place can be classified as rotational, vibrational, or electronic. 
While for the latter, electronic excitation is responsible for reaching the ionization continuum, in the two former cases it is an excess of nuclear rotational or vibrational energy that brings the system above the ionization threshold. 
The autoionization process itself can then be viewed as a nonadiabatic transition (internal conversion) between the initial rotationally or vibrationally excited $N$-electron molecule and the final system of $N-1$-electron molecule and free electron, where the $N-1$-electron molecule bears a reduced internal energy. 
Vibrational autoionization has been identified over the last decades in a variety of molecules, ranging from small anionic diatomics,\cite{lineberger83,lineberger85} Rydberg-excited states of neutral molecules\cite{pratt2005} to dipole- or quadrupole-bound states of polyatomic anions\cite{johnson00,scheier06,weber10,wang13,wang14,verlet16,wang17,weber19,verlet20,verlet21,wang22} and has been utilized to measure highly resolved photodetachment spectra. In some cases, vibrational autoionization from valence states has been observed as well.\cite{gerardi,devineJPCL,verlet19}
Real-time access to the dynamics of such processes could be recently gained via pump-probe experiments.\cite{kang20,kang21,verlet21}

Theoretical considerations of vibrational autoionization have been originally outlined by Berry\cite{berry66} and later formulated for anions by Simons, who established propensity rules for such transitions for model cases\cite{proprulessimons} and computed autoionization rates for several small molecules using time-independent\cite{simons84} as well as time-dependent pictures.\cite{simons99}
Going beyond the calculation of rates and fully simulating the real-time dynamics of such processes for complex molecules is highly desirable, but a completely quantum mechanical treatment, including the full dimensional nuclear motion and the description of electron scattering states, is computationally prohibitive. 
However, given the great success of mixed quantum-classical approaches to describe the bound-state nonadiabatic dynamics\cite{werner08,thiel2009, fish, Persico2014, worth2015, martinez18,barbatti2018} as well as time-resolved spectroscopic observables\cite{humeniuk2013,2ma,anjafeature} of a large variety of molecules, it suggests itself to consider such an ansatz as well for vibration-induced autoionization. 
Recently, the dynamics of electronically metastable anion states has been addressed by combining classical trajectory simulations with quantum mechanically calculated ionization probabilities based on the width of the electronic resonance.\cite{barbatti19,jagau22}
In the present work, we will introduce a novel methodology for the dynamics of vibrationally metastable anions which is based on Tully's trajectory surface hopping,\cite{tully} treating the nuclear motion classically, while retaining the quantum mechanical description of the electronic system. 
The energy exchange between the electronic and nuclear subsystems will be described by nonadiabatic transitions between the bound and continuum electronic states, accompanied by an associated change of the classical vibrational energy of the molecule.

After presenting the theory we will illustrate our method on the example of the vinylidene anion, C$_2$H$^-_2$. \textit{Neutral} vinylidene is a high-energy isomer of the well-known acetylene molecule, HCCH, to which it readily isomerizes.\cite{isomerization1,isomerization2,isomerization3} 
As an anion, however, vinylidene is stable on the time scale of seconds,\cite{andersen00} while the acetylene anion is electronically unbound.\cite{isomerization3,acetylenanion} 
Vinylidene anions can be produced in the gas phase e.g. by injection of electrons in a precursor gas mixture containing ethylene and N$_2$O, where the reaction proceeds via intermediately formed O$^-$.\cite{gerardi} 
Using photodetachment and photoelectron spectroscopy techniques, the vinylidene anion has been utilized to gain information about the electronic and vibrational states of neutral vinylidene.\cite{burnett83,isomerization3,neumark16} 
In this way, it could be established that neutral vinylidene represents a local minimum on the C$_2$H$_2$ potential energy surface with several clearly assignable vibrational states, while with higher vibrational energy, the distinction between the vinylidene and acetylene isomers gets lost. Therefore, as shown by quantum dynamics simulations, isomerization to acetylene readily occurs on a sub-picosecond time scale upon vibrational excitation, while for the vibrational ground state much larger lifetimes of several hundred picoseconds can be expected.\cite{qdvinylidene1,bowman03,qdvinylidene2} In fact, experiments employing Coulomb explosion imaging provided even evidence for the presence of vinylidene on the much longer time scale of several microseconds after its initial generation.\cite{levin98} With the help of classical molecular dynamics simulations, it could be shown that this finding is due to frequent forth- and back-formation of vinylidene after initial isomerization provided the vibrational energy is sufficient to overcome the isomerization barrier.\cite{carter01}

With regard to the ionization process itself, photoelectron spectroscopy studies have enabled the determination of vinylidene's adiabatic electron affinity (AEA), with the current most precise value being 0.4866(8) eV (3935 cm$^{-1}$).\cite{devineScience} 
Employing vibrational predissociation spectroscopy of Ar-tagged vinylidene anions, the vibrational structure of the anionic ground state has been investigated, revealing prominent spectral features around 2600 and 4000 cm$^{-1}$.\cite{gerardi} 
As the latter value lies above the AEA, it is also visible as a resonance in photodetachment spectroscopy, indicating that the respective vibrational states couple to the ionization continuum. Interestingly, also the lower-energy band around 2600 cm$^{-1}$ appears prominently in photodetachment, although under the experimental conditions used in Ref. \onlinecite{gerardi} one-photon ionization is dominant and thus direct detachment should be improbable. This finding could be attributed to the presence of molecules that are initially occupying excited vibrational states due to thermal energy and are further photoexcited above the AEA, followed by vibrational autoionization. The photoelectron spectra resulting from such detachment processes allow for conclusions on the vibrational structure of the neutral species as well as on the initially populated anion vibrational states. In addition, recent experiments employing slow electron velocity imaging (SEVI) photoelectron spectroscopy provided highly resolved spectra (<10 cm$^{-1}$) that allowed one to disentangle further structures in the photoelectron spectra obtained from the vibrational resonances just above the ionization threshold.\cite{devineJPCL} Specifically, two types of spectral features were found: (i) peaks with electron kinetic energies shifting proportionally to the incident photon energy, as is expected from a direct photodetachment process, and (ii) peaks of constant electron kinetic energy over a range of photon energies, covering a region up to 100 cm$^{-1}$. This finding can be explained by intermediate excitation of rovibrational states of the anion that are resonant to the photon energy and decay via autoionization to the neutral species. In this process the rotational quantum numbers remain unchanged, hence the occurrence of constant-energy photoelectrons.
These findings provided direct evidence for vibration-induced autoionization processes following the excitation of anion vibrational states above the ionization threshold. What has remained undisclosed until now, however, are their time scales. 
The present paper aims to shed light on this dynamical process by simulating directly the autoionization dynamics in full dimensionality.

Our paper is organized as follows: In section \ref{sec:Theory} the theoretical methodology will be presented and in section \ref{sec:Comp} the computational details are given. This is followed by the results and discussion provided in section \ref{sec:Results}. Finally, conclusion and outlook are given in section \ref{sec:Concl}.

\section{\label{sec:Theory}Theoretical Approach}
In the frame of the Born-Oppenheimer approximation, a vibrational resonance can be described by a product of a vibrational and an electronic wavefunction, $\chi_{vib}(\varepsilon)\Phi_{el}^{N}(E)$, with vibrational energy $\varepsilon$ and electronic energy $E$, where the total energy $E+\varepsilon$ exceeds the ionization threshold while the electronic portion is still below it. 
The autoionization process can then be viewed as the internal conversion to an isoenergetic state $\chi_{vib}(\varepsilon')\Phi^N_{el}(E')$ in which part of the vibrational energy has been transformed into electronic energy, leading to an electronic state with increased energy $E'$ that is unbound with respect to single-electron loss and can be described as an antisymmetrized product of an $N-1$ electron bound state and a free electron continuum state with wave vector $\textbf{k}$, $\Phi^N_{el}(E')={\cal A} (\Phi^{N-1}_{el}\psi(\textbf{k}))$.

Since we aim to establish a method that is applicable for complex molecules, instead of a fully quantum mechanical description of the autoionization dynamics a mixed quantum-classical picture is desirable, in which only the quantum nature of the electronic part is retained while the nuclear motion is described classically. For nonadiabatic processes between bound electronic states, the surface hopping method\cite{tully} has proven to be a very versatile approach.  In this framework, the nuclear degrees of freedom are propagated classically by solving Newton's equations of motion, 
\begin{eqnarray}
M\ddot{\textbf{R}}=-\nabla_\textbf{R}E_i(\textbf{R}),\label{newton}
\end{eqnarray}
for an ensemble of initial conditions, thereby giving rise to nuclear trajectories $\textbf{R}(t)$ moving on an electronic potential energy surface $E_i(\textbf{R})$. The quantity $M$ denotes a diagonal matrix containing the nuclear masses. In parallel, along each trajectory an electronic time-dependent Schr\"{o}dinger equation is solved, which most generally reads
\begin{eqnarray}
i\hbar \frac{d}{dt}\Psi(\textbf{r},t;\textbf{R}[t])=\hat{H}_{el}\Psi(\textbf{r},t;\textbf{R}[t]),\label{el_schroedinger}
\end{eqnarray}
with $\hat{H}_{el}$ being the electronic Hamiltonian of the system. 
The electronic wavefunction can be expanded with respect to a set of orthogonal basis states as 
\begin{eqnarray}
\Psi(\textbf{r},t;\textbf{R}[t])=\sum_i c_i(t) \Phi_i\big(\textbf{r};\textbf{R}[t]\big),\label{adiab_expansion}
\end{eqnarray}
leading to the following set of coupled differential equations for the coefficients $c_i$:
\begin{eqnarray}
i\hbar\dot{c}_i(t)=\sum_j \left [ H_{ij}(\textbf{R}[t])-i\hbar D_{ij} (\textbf{R}[t])\right ]c_j(t),
\label{schroedinger}
\end{eqnarray}
where $H_{ij}=\langle  \Phi_i|\hat{H}_{el}| \Phi_j\rangle$ denotes the matrix elements of the electronic Hamiltonian. The $D_{ij}$ represent the nonadiabatic couplings that arise from the parametric dependence of the wavefunction on the nuclear trajectory and can be written as
\begin{eqnarray}
D_{ij}&=&\langle \Phi_i|\dot{\Phi}_j\rangle=\dot{\textbf{R}}\cdot \langle\Phi_i|\nabla_R|\Phi_j\rangle
\end{eqnarray}
where the last expression makes clear the dependence on the nuclear velocities $\dot{\textbf{R}}$. Thus, the coupling between electronic states is mediated by the nuclear motion. The time-dependent coefficients $c_i(t)$ are employed in the surface hopping approach to devise probabilities for each nuclear trajectory to switch its electronic state, and the time-dependent properties of the nonadiabatically evolving system are obtained by averaging the quantities of interest over the entire ensemble that typically contains up to hundreds of trajectories. 

In the following, we will develop a surface-hopping methodology for the description of molecular autoionization dynamics.
For a system that can be ionized, the expansion of the electronic state given in Eq. (\ref{adiab_expansion}) has to be extended by the set of continuum eigenstates:
\begin{eqnarray}
\Psi\big(\textbf{r},\textbf{R}[t],t\big) = \sum_m c_m(t) \Phi_m\big(\textbf{r},\textbf{R}[t]\big)\nonumber
\\+\sum_n\int\! d^3\textbf{k}\;\tilde{c}_n(\textbf{k},t)\tilde{\Phi}_n(\textbf{k},\textbf{r},\textbf{R}[t]),
\end{eqnarray}
where the first sum includes the $N$-electron bound states of the molecule, while the second sum and integral encompass the set of singly ionized states characterized by the discrete quantum number $n$ of the bound $N-1$-electron system and the continuously varying wavevector $\textbf{k}$ of the free electron. Bound and continuum eigenstates are mutually orthogonal in the sense
\begin{align}
\langle\Phi_m|\Phi_{m'}\rangle&=\delta_{mm'}&&\textrm{(bound-bound)}\\
\langle\Phi_m|\tilde{\Phi}_n(\textbf{k})\rangle&=0&&\textrm{(bound-continuum)}\\
\langle\tilde{\Phi}_n(\textbf{k})|\tilde{\Phi}_{n'}(\textbf{k}')\rangle&=\delta_{nn'}\delta(\textbf{k}-\textbf{k}').&&\textrm{(continuum-continuum)}\nonumber\\\label{k-norm}
\end{align}
In the following, we will specifically consider the ionization of a negatively charged molecule, thus our $N$-electron system is an anion (a), the $N-1$-electron system is neutral (n). In order to simulate the dynamics of the autoionization process on similar grounds as bound state nonadiabatic dynamics, several approximations have to be introduced. 

\subsection{\label{discretized}Discretized continuum approximation for ionized states}
We discretize the set of continuum states as 
\begin{eqnarray}
&&\int d^3\textbf{k}\,\tilde{c}_n(\textbf{k},t)\tilde{\Phi}_n(\textbf{k},\textbf{r},\textbf{R}[t])\\
&\approx& \sum_i (\Delta V_k)^\frac{1}{2} \tilde{c}_n(\textbf{k}_i,t) (\Delta V_k)^\frac{1}{2}\tilde{\Phi}_n(\textbf{k}_i,\textbf{r},\textbf{R}[t])\\
&\approx&  \sum_i c_n(\textbf{k}_i,t)  \Phi_n(\textbf{k}_i,\textbf{r},\textbf{R}[t])
\end{eqnarray}
where $\Delta V_k$ is the approximate volume element in k-space, and the continuum and discretized versions of coefficients and wave functions are related according to $c_n(\textbf{k}_i,t)=(\Delta V_k)^\frac{1}{2} \tilde{c}_n(\textbf{k}_i,t)$ and $\Phi_n=(\Delta V_k)^\frac{1}{2}\tilde{\Phi}_n$. 

\begin{figure}[b]
\includegraphics[width=0.8\columnwidth]{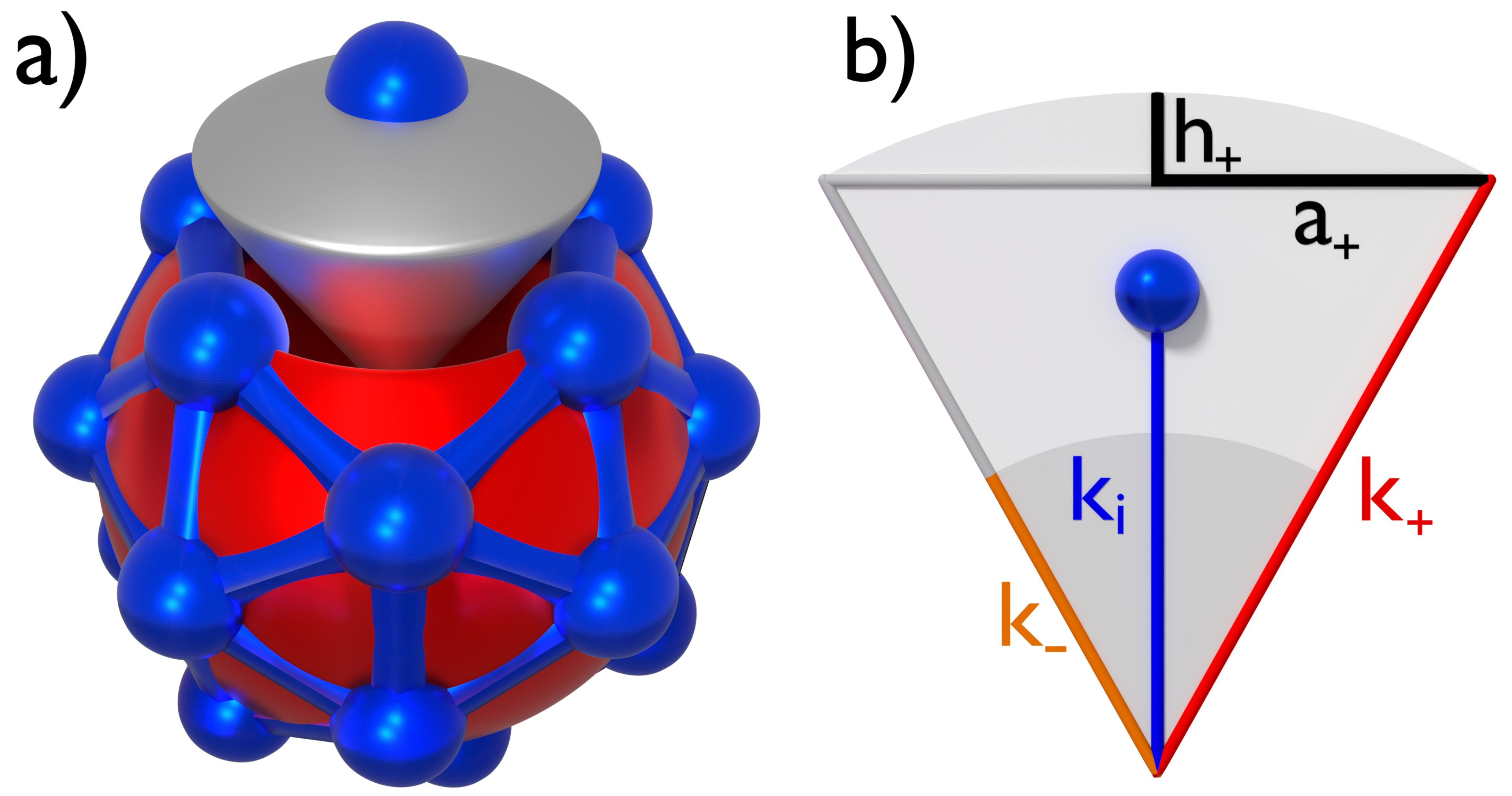}
\caption{\label{fig:snubcube} (a) Snub cube arrangement of 24 points on a spherical surface. A selected spherical sector (grey) is slightly lifted for illustrative purposes. (b) Cut through the spherical sector with assignment of several k-space distances discussed in the text.}
\end{figure}

For the discretization of k-space, different approaches can be employed. Most simply, the Cartesian components of the k-vector can be discretized evenly, leading to the approximation of the volume element in k-space as
\begin{eqnarray}
d^3\textbf{k}\approx\Delta V_k = \Delta k_x\Delta k_y\Delta k_z.
\end{eqnarray}
While being conceptually straightforward (especially if the same spacing is used for all three spatial directions), this approximation may be considered not optimal if one aims at analyzing the free electrons in terms of directions and kinetic energies (the latter depending on the length of the k-vector).  
Therefore, we employed an alternative discretization scheme where the absolute values of the k-vectors were discretized such that a given energy region was evenly covered. For each energy value, the orientations of the k-vector were discretized by distributing points approximately uniformly on the corresponding spherical surface. This is equivalent to the well-known Thomson problem of finding the optimal placement of electrical charges on a sphere so as to minimize their repulsion energy. For the case of 24 points, the optimal distribution is exactly known and results in a snub cube as depicted in Fig. \ref{fig:snubcube}a. For an arbitrary number of surface points, the optimal distribution can be approximately determined using, e.g., the Fibonacci sphere algorithm.\cite{fibonacci} To determine the volume element covered by each point $\textbf{k}_i$, we consider the volume difference of two spherical sectors with k-space radii $k_+$ and $k_-$ (cf. Fig. \ref{fig:snubcube}b) corresponding to the energies $E_\pm=E(\textbf{k}_i)\pm \frac{\Delta E}{2}$, where $\Delta E$ is the fixed discretization width for the energies, such that
\begin{eqnarray}
k_\pm = \sqrt{k_i^2\pm \frac{m_e\Delta E}{\hbar^2}}
\end{eqnarray}
and hence
\begin{equation}
d^3\textbf{k} 
\approx\Delta V_k=
V_+ - V_- = \frac{2\pi}{3} (k_+^2h_+ - k_-^2h_-)
\end{equation}
The height $h_\pm$ of the spherical cap can be expressed using the cap radius $a_\pm$ and the sphere radius $k_\pm$:
\begin{eqnarray}
h_\pm = k_\pm - \sqrt{k_\pm^2-a_\pm^2},
\end{eqnarray}
leading to
\begin{eqnarray}
\Delta V_k 
&=    
\frac{2\pi}{3} \Big(k_+^2\big(k_+ - \sqrt{k_+^2-a_+^2}\big) - k_-^2\big(k_- - \sqrt{k_-^2 - a_-^2}\big)\Big)\nonumber\\
\end{eqnarray}
The cap radius $a_\pm$ is linearly dependent on the k-space radius $k_\pm$, $a_\pm = \tilde{a} k_\pm$, with $\tilde{a}$ being independent of the radius, which results in
\begin{eqnarray}
\Delta V_k 
=\frac{2\pi}{3} \big(1-\sqrt{1-\tilde{a}^2}\big)\big(k_+^3 - k_-^3\big) = c \big(k_+^3 - k_-^3\big).
\end{eqnarray}
The diameter $2 a_\pm$ of the spherical cap is taken as the average distance between a specific point on the sphere and the six points surrounding it. For the snub cube, this corresponds to a universal $\tilde{a}$ value of $\approx 0.39779$. The sum of spherical cap surfaces obtained in this way results in a surface area deviating from the actual spherical surface by less than 3 \% for the Fibonacci algorithm and only 1 \% for a snub cube, and in a sphere volume of similar accuracy, therefore justifying the approximation. 

As a next step, the discretized continuum state expansion obtained this way is inserted into the time-dependent Schr\"{o}dinger equation (\ref{el_schroedinger}) to derive the equations of motion for the electronic degrees of freedom, as detailed below.\smallskip

\subsection{\label{discretizedTDSE}Time-dependent Schr\"{o}dinger equation in the discretized continuum approximation}
After insertion into the time-dependent Schr\"{o}dinger equation (\ref{el_schroedinger}), the discretized continuum state expansion is projected on the electronic basis states, resulting in a set of coupled equations of motion for the bound and continuum state coefficients completely analogous to Eq. (\ref{schroedinger}). In the expressions below, $m,m'$ denote the bound electronic states of the anion, $n,n'$ the bound electronic states of the neutral molecule, and $i$ is the index counting the discretized scattering states of the detached electron:
\begin{widetext}

\begin{align}
\textrm{(bound)}&&
i\hbar\dot{c}_m(t)
=& 
\sum_{\substack{m',\textrm {anion}}} 
\Big[
H_{mm'}(\textbf{R}[t])
-i\hbar 
D_{mm'}(\textbf{R}[t])
\Big]
c_{m'}(t)
+
\sum_{\substack{n',\textrm {neutral}}}
\sum_{i,\textrm{free}} 
\Big[
H_{mn'}(\textbf{k}_i,\textbf{R}[t])
-i\hbar 
D_{mn'}(\textbf{k}_i,\textbf{R}[t])
\Big]
c_{n'}(\textbf{k}_i,t)
\label{schroedinger2}\\
\textrm{(continuum)}&&
i\hbar\dot{c}_n(\textbf{k}_i,t)
=&
\sum_{\substack{m',\textrm {anion}}} 
\Big{[}
H_{nm'}(\textbf{k}_i,\textbf{R}[t])
-i\hbar 
D_{nm'}(\textbf{k}_i,\textbf{R}[t])
\Big{]}
c_{m'}(t)
\label{schroedinger3}
\end{align}

\end{widetext}
with the diabatic and nonadiabatic couplings between two bound anion states,
\begin{eqnarray}
H_{mm'}&=&\langle \Phi_m|\hat{H}|\Phi_{m'}\rangle\\
D_{mm'}&=&\langle \Phi_m|\dot{\Phi}_{m'}\rangle,\label{dij}
\end{eqnarray}
and between a bound and a discretized continuum state,
\begin{eqnarray}
H_{nm'}(\textbf{k}_i)&=&(\Delta V_k)^\frac{1}{2}\langle\tilde{\Phi}_{n}(\textbf{k}_i) |\hat{H}|\Phi_{m'}\rangle\\
D_{nm'}(\textbf{k}_i)&=&\langle\dot{\Phi}_{n}(\textbf{k}_i) |\Phi_{m'}\rangle=(\Delta V_k)^\frac{1}{2}\langle \dot{\tilde{\Phi}}_{n}(\textbf{k}_i)|\Phi_{m'}\rangle.\label{dik}
\end{eqnarray}
The coupling among the discretized continuum states has been neglected in the above equations.

\subsection{\label{planewave}Plane-wave approximation for continuum states}
As we treat the electron detachment from anions, the continuum states $\tilde{\Phi}_n(\textbf{k}_i)$ correspond to an antisymmetrized linear combination of a bound state of the neutral molecule and a molecular scattering state of the free electron,
\begin{eqnarray}
\tilde{\Phi}_n(\textbf{k}_i)={\cal A} \left (\Phi^{\textrm{(n)}}_n\cdot\psi(\textbf{k}_i)\right )
\end{eqnarray}
The simplest approximation to the scattering continuum, with asymptotic wave vector $\textbf{k}_i$, is provided in such a case by using plane waves,
\begin{eqnarray}
\psi(\textbf{k}_i)\approx {\cal N} \textrm{e}^{i\textbf{k}_i\cdot \textbf{r}}.
\end{eqnarray} 
We set ${\cal N}=(2\pi)^{-3/2}$ such that for $\tilde{\Phi}_n(\textbf{k}_i)$ the normalization condition (\ref{k-norm}) is fulfilled. For the \textit{discretized} state $\Phi_n(\textbf{k}_i)$, this corresponds to normalization within a spatial box of length $L=\frac{2\pi}{\Delta k}$, such that $\langle \Phi_n (\textbf{k}_i)|\Phi_{n'} (\textbf{k}_j)\rangle_{\textrm{box}}=\delta_{nn'}\delta_{ij}$.
The free electron state obtained this way bears no dependence on the molecular structure, which is a strong simplification. In order to include this dependence at least to a certain extent, we consider plane waves orthogonalized with respect to the occupied molecular orbitals (MOs) $ \phi_m(\textbf{r},\textbf{R}[t])$ of the anion, 
\begin{eqnarray}
\tilde{\psi}(\textbf{k}_i)&=& (2\pi)^{-3/2}\left ( \textrm{e}^{i\textbf{k}_i\cdot \textbf{r}} - \sum_m^{\mathrm{occ}} \langle \phi_m| \textrm{e}^{i\textbf{k}_i\cdot \textbf{r}}\rangle \phi_m\right )\\
&=&\psi(\textbf{k}_i)-\sum_m^{\mathrm{occ}} \langle \phi_m| \psi(\textbf{k}_i)\rangle\, \phi_m.\label{orth_pw}
\end{eqnarray}
The energy of such a discretized continuum state can be approximated as
\begin{eqnarray}
E_n(\textbf{k}_i)&=&E^{\textrm{(n)}}_n+\frac{k^2_i}{2m_e}
\end{eqnarray}
with $E^{\textrm{(n)}}_n$ as the electronic energy of the neutral molecule and $\frac{k^2_i}{2m_e}$ as the asymptotic kinetic energy of the free electron.

Employing this orthogonalized plane wave approximation, the matrix elements for the electronic couplings between the bound and continuum states can now be constructed as will be outlined in the following section.

\subsection{\label{nonad}Electronic coupling}
The electronic wavefunctions employed in our approach are determined using separate quantum chemical calculations for the anionic and the neutral molecule. Therefore, they are approximations to the bound adiabatic eigenfunctions of the electronic Hamiltonian for either $N$ or $N-1$ electrons, and thus only the diagonal matrix elements of the bound-state electronic Hamiltonians are non-zero. However, due to the neglect of the \textit{unbound} eigenfunctions in the numerical solution of the electronic Schr\"{o}dinger equation, there is also a non-zero diabatic coupling between the bound and continuum $N$-electron states present. Additionally, dynamical nonadiabatic couplings occur in the time-dependent Schr\"{o}dinger equation in complete analogy to the case of bound-state dynamics. In the following, we outline our approach to calculate these couplings in an approximate way.
\medskip
\subsubsection{Diabatic coupling}
In the following, we consider ionization transitions between the electronic ground states of both the anionic and the neutral species. As a shorthand notation, we label by $\Phi_0$ the electronic ground state of the anion, and by $\Phi_i$ the discretized continuum state where the neutral molecular core is also in the ground state and the free electron has wave vector $\textbf{k}_i$. Thus, we can write the diabatic matrix element between these states as
\begin{eqnarray}
H_{00}(\textbf{k}_i)\equiv\langle \Phi_i|\hat{H}| \Phi_0\rangle\equiv(\Delta V_k)^\frac{1}{2}\, V^{\mathrm{dia}}_{i0}(\textbf{k}_i).
\end{eqnarray}
We assume the two ground-state wavefunctions $\Phi_0$ and $\Phi_i$ to be represented by single Slater determinants. For the bound anionic system, these are as usual constructed from the occupied anion molecular orbitals (MOs) $\phi_p$. For the ionized system, the Slater determinant is formed from the orthogonalized plane wave $\tilde{\psi}(\textbf{k}_i)$ and the occupied neutral MOs $\chi_n$. The two wavefunctions are thus constructed by two mutually non-orthogonal sets of MOs. The derivation of matrix elements between such wavefunctions has been performed by first by L\"{o}wdin.\cite{lowdin55} Along these lines, we show in Appendix \ref{App_A} that expansion with respect to the MOs leads to the following expression for the diabatic coupling:
\begin{eqnarray}
&&V_{i0}^{\mathrm{dia}}(\textbf{k}_i)=\sum_p\langle \tilde{\psi}|\hat{h}|\phi_p\rangle(-1)^{p+1} \mathrm{det}\,\mathbf{S}_{i,p}\\
&&+\sum_n \langle \tilde{\psi}\chi_n|\hat{v}|\sum_{q,p<q}\left ( \phi_p\phi_q-\phi_q\phi_p\right )\rangle \cdot (-1)^{n+p+q-1} \mathrm{det}\,\mathbf{S}_{in,pq}, \nonumber
\end{eqnarray}
where $\hat{h}$ and $\hat{v}$ denote the usual one- and two-electron parts of the electronic Hamiltonian, $\mathrm{det}\,\mathbf{S}_{i,p}$ is the minor determinant obtained from the overlap matrix 
\begin{eqnarray}
\mathbf{S}=\begin{pmatrix}
\langle \tilde{\psi}|\phi_1\rangle & ... & \langle \tilde{\psi}|\phi_p\rangle&...&\langle \tilde{\psi}|\phi_N\rangle\\
\langle \chi_1|\phi_1\rangle & ... & \langle \chi_1|\phi_p\rangle&...&\langle \chi_1|\phi_N\rangle\\
... & ... & ...&...&...\\
\langle \chi_n|\phi_1\rangle & ... & \langle \chi_n|\phi_p\rangle&...&\langle \chi_n|\phi_N\rangle\\
... & ... & ...&...&...\\
\langle\chi_{N-1}|\phi_1\rangle&...&\langle\chi_{N-1}|\phi_p\rangle&...&\langle\chi_{N-1}|\phi_N\rangle
\end{pmatrix},
\end{eqnarray}
between bound- and continuum state orbitals by deleting the row of the plane wave $\tilde{\psi}(\textbf{k}_i)$ and the column of the anion MO $\phi_p$, while $\mathrm{det}\,\mathbf{S}_{in,pq}$ denotes the minor determinant obtained by deleting the rows of the plane wave and the neutral MO $\chi_n$ as well as the columns of the anion MOs $\phi_p$ and $\phi_q$.

By expanding the neutral molecule's MOs with respect to the occupied and virtual anion MOs it can be shown that the one-electron contributions and those two-electron contributions associated with the occupied anion MOs vanish, as detailed in Appendix \ref{App_A}. Therefore, the final MO form of the diabatic coupling only includes electron-electron interactions between the plane wave and the virtual anion MOs on the one hand and the occupied anion MOs on the other hand:
\begin{widetext}
\begin{eqnarray}
V_{i0}^{\mathrm{dia}}(\textbf{k}_i)=\sum_n^{\mathrm{occ}} \sum_u^{\mathrm{virt}} \langle \chi_n|\phi_u\rangle \langle \tilde{\psi}\phi_u|\hat{v}|\sum_{q,p<q}^{\mathrm{occ}}\left ( \phi_p\phi_q-\phi_q\phi_p\right )\rangle \cdot (-1)^{n+p+q-1} \mathrm{det}\,\mathbf{S}_{in,pq},\label{diabatic_twoel}
\end{eqnarray}
\end{widetext}
where $\langle \chi_n|\phi_u\rangle$ denotes the overlap integral between an occupied neutral and a virtual anion MO. Inserting the definition of the orthogonalized plane waves and expansion of the MOs with respect to the atomic orbital (AO) basis, as described in detail in Appendix \ref{App_dia_AO}, finally leads to working equations for the diabatic coupling that involve overlap and electron-electron repulsion integrals between basis functions and/or plane waves: 
\begin{alignat}{3}
V^{\mathrm{dia}}_{i0}(\textbf{k}_i)
&=&
\sum_{\lambda\mu\nu}
    \Bigg[&
        A_{\lambda\mu\nu}
        \Big(
            \langle
                \mathbf{k}_i \lambda || \mu \nu
            \rangle
            -\sum_{\sigma}
                B_{\sigma}
                \langle
                    \sigma \lambda || \mu \nu
                \rangle
        \Big) + \nonumber\\
        &&&
        \bar{A}_{\lambda\mu\nu}
        \Big(
            \langle
                \mathbf{k}_i \lambda | \mu \nu
            \rangle
            -\sum_{\sigma}
                B_{\sigma}
                \langle
                    \sigma \lambda | \mu \nu
                \rangle
        \Big)
    \Bigg]\label{vdia_ao_main}
\end{alignat}
where the Greek indices denote AO basis functions and $\langle ab||cd\rangle=\langle ab|cd\rangle-\langle ab|dc\rangle$ is an antisymmetrized electron-electron repulsion integral. The prefactors $A_{\lambda\mu\nu}$, $\bar{A}_{\lambda\mu\nu}$ and $B_{\sigma}$ are computed from AO expansion coefficients and overlap integrals as detailed in Appendix B, Eqn. (\ref{Afactor})-(\ref{Bfactor}). As the appropriate basis functions for molecular calculations we employ Cartesian Gaussian functions of the form
\begin{eqnarray}
\varphi_\nu(\mathbf{r})=\prod_{j=x,y,z}(r_j-A_{\nu, j})^{n_{\nu,j}}\, \textrm{e}^{-\alpha_\nu(\textbf{r}-\mathbf{A}_\nu)^2},
\end{eqnarray} 
where $\mathbf{A}_\nu$ denotes the center of the Gaussian (usually an atomic position), and the $n_{\nu,j}$ are angular momentum quantum numbers.

The calculation of the diabatic coupling according to Eq. (\ref{vdia_ao_main}) thus involves four types of integrals:
\paragraph{Overlap integrals between MOs} These can be reduced to overlaps between Gaussian basis functions which are analytically calculated according to Ref. \onlinecite{taketa66}
\paragraph{Overlap integrals between a Gaussian and a plane wave} These correspond to inverse Fourier transforms of the basis functions and are analytically calculated as
\begin{eqnarray}
\langle \textbf{k}|\nu\rangle&=&\frac{1}{(2\pi)^{3/2}}\int d^3\mathbf{r}\,\textrm{e}^{-i\textbf{k}\cdot \textbf{r}}\varphi_\nu(\textbf{r})\\
&=&\left(\frac{1}{2\alpha_\nu}\right)^{\frac{3}{2}} \mathrm{e}^{-i\textbf{k}\cdot\textbf{A}_\nu -\frac{k^2}{4\alpha_\nu }}  \nonumber\\
&&\times \prod_{j=x,y,z} \left(\frac{-i}{2\sqrt{\alpha_\nu}}\right)^{n_{\nu,j}}H_{n_{\nu,j}}\left(\frac{k_j}{2\sqrt{\alpha_\nu }}\right)\label{ft_gauss},
\end{eqnarray}
where the $H_{n_{\nu,j}}$ are the Hermite polynomials of order $n_{\nu,j}$.
\paragraph{Electron-electron repulsion integrals between Gaussian basis functions} For these integrals, efficient analytical expressions have been derived in the literature and used in quantum chemical programs. In the present contribution, we employ the Rys quadrature method\cite{dupuis76,rys83} as implemented in the \texttt{libcint}\cite{libcint} library.
\paragraph{Electron-electron repulsion integrals between a plane wave and Gaussian basis functions} Analytical formulae for these integrals have been reported,\cite{watson79,colle87} but are not commonly available in molecular quantum chemistry codes. As an efficient alternative, we instead employ approximate formulae based on the observation that, for most of the Gaussian functions in our basis set, the plane waves under consideration do not strongly change within the width of the Gaussian. Therefore, in the integral $\langle \mathbf{k}_i\lambda|\mu\nu\rangle$ it is reasonable to replace the plane wave by the first terms of its Taylor expansion around the center $\textbf{R}_\mu$ of the Gaussian $\varphi_\mu(\textbf{r}_1)$ according to
\begin{eqnarray}
\textrm{e}^{-i\textbf{k}_i\cdot \textbf{r}_1}&\approx&\textrm{e}^{-i\textbf{k}_i\cdot \textbf{R}_\mu}\left [ 1 - i\textbf{k}_i\cdot \left ( \textbf{r}_1-\textbf{R}_\mu\right )\right ] \label{exp_appr}
\end{eqnarray}
In this way, the integrals reduce to common Gaussian three-center two-electron integrals, which are also used as implemented in \texttt{libcint}. Systematic tests of the approximation are summarized in Table \ref{tab:table0}. We find good accuracy for small values of the plane wave energy, with discrepancies increasing with growing energies. Up to 0.5 eV, which represents the range most relevant for the present study, the errors are still moderate.

\begin{table}[b]
\caption{\label{tab:table0}%
Average values (in $E_H$) and errors (in \%) of hybrid Gaussian-plane wave electron repulsion integrals $\langle \mathbf{k}_i\lambda|\mu\nu\rangle$ for vinylidene employing the d-aug-cc-pVDZ basis set. The molecular structure has been optimized at the DFT/$\omega$B97XD/d-aug-cc-pVDZ level. $I_{ex}$ denotes the exact integral, $I_{ap}$ the approximate value according to Eq. (\ref{exp_appr}). For each plane wave energy $E$, the average has been taken over all distinct integrals provided by the basis set as well as over 24 different k-vectors corresponding to the direction vectors of the vertices of a snub cube.
}
\begin{ruledtabular}
\begin{tabular}{ccccc}
E/$eV$  & $\langle|I_{ex}|\rangle$ &  $ \frac{\langle|I_{ex}-I_{ap}|\rangle}{\langle|I_{ex}|\rangle} $ & $ \left\langle \frac{|I_{ex}-I_{ap}|}{|I_{ex}|} \right\rangle^{[a]}$ \\
\colrule
0.0015  & 0.249 &  $1\cdot 10^{-3}$ & 13.0 \\
0.1  & 0.276 & $9\cdot 10^{-2}$ & 19.7 \\
0.5 & 0.275 & 0.5 & 31.3  
\end{tabular}
\end{ruledtabular}
$^{[a]}$ Average computed for all integrals with $|I_{ex}|>10^{-16}\,E_H$.
\end{table}

\subsubsection{Nonadiabatic coupling}
The molecular wavefunctions $\Phi_0$ and $\Phi_i$ of the bound anion and the continuum state are not strictly diabatic but still bear a dependence on the nuclear geometry. Therefore, in addition to the diabatic coupling, there is also residual nonadiabatic coupling 
\begin{equation}
D_{i0}(\textbf{k}_i)=\langle\Phi_i(t)|\frac{d}{dt}{\Phi}_0(t)\rangle  
\end{equation}
present, which we calculate employing a finite-difference approximation for the time derivative similar to the procedure presented in Ref. \onlinecite{mitric08_BAN}:
\begin{eqnarray}
D_{i0} &\approx& \frac{1}{2\Delta t} \Big( \langle\Phi_i(t)|\Phi_0(\tau)\rangle - \langle\Phi_i(\tau)|\Phi_0(t)\rangle \Big)\nonumber
\end{eqnarray}
with $\tau\equiv t+\Delta t$. The quantity $\langle\Phi_i(t)|\Phi_0(\tau)\rangle$ reduces to a one-electron integral of the form 
\begin{eqnarray}
(\sqrt{N})^{-1}\langle \tilde{\psi}(\textbf{k}_i,t)|\psi^D(t,\tau)\rangle,\label{pw_dyson}
\end{eqnarray}
involving the function 
\begin{eqnarray}
\psi^D(t,\tau)=\sqrt{N}\langle \Phi^{N-1}_0(t)|\Phi^{N}_0(\tau)\rangle_{1...N-1}.
\end{eqnarray}
which can be regarded as an analog to a molecular Dyson orbital, but with the $N$- and $N-1$-electron wavefunctions taken at different time steps, i.e., different molecular geometries. Inserting the definition of the orthogonalized plane wave $\tilde{\psi}(\textbf{k}_i,t)$, Eq. (\ref{orth_pw}), into expression (\ref{pw_dyson}) leads to
\begin{eqnarray}
\langle\Phi_i(t)|\Phi_0(\tau)\rangle&=&(\sqrt{N})^{-1}\big [\langle \psi(\textbf{k}_i)|\psi^D(t,\tau)\rangle\big.\\
&&\big.-\sum_n \langle \psi(\textbf{k}_i)|\phi_n(t)\rangle\langle\phi_n(t)|\psi^D(t,\tau)\rangle \big].\nonumber
\end{eqnarray}

For the nonadiabatic couplings, this gives rise to the final expression
\begin{widetext}
\begin{eqnarray}
D_{i0}(\textbf{k}_i) = \frac{(\Delta V_k)^\frac{1}{2}}{2\sqrt{N}\Delta t} &&\Big{[} \left \langle \psi(\textbf{k}_i)|\psi^D(t,\tau)\right\rangle-\sum_n \left\langle \psi(\textbf{k}_i)|\phi_n(t)\right\rangle\,\left\langle\phi_n(t)|\psi^D(t,\tau)\right\rangle \Big.\nonumber\\
&&\Big.-\left\langle \psi(\textbf{k}_i)|\psi^D(\tau,t)\right\rangle+\sum_n \left\langle \psi(\textbf{k}_i)|\phi_n(\tau)\right\rangle\,\left\langle\phi_n(\tau)|\psi^D(\tau,t)\right\rangle \Big{]} \label{nonad_pw_mo}
\end{eqnarray}
\end{widetext}

The Dyson orbitals $\psi^D$ are constructed following the procedure outlined by Humeniuk et al.,\cite{humeniuk2013} which eventually leads to their representation as linear combinations of atomic basis functions,
\begin{eqnarray}
\psi^D(t,\tau)=\sum_\nu c_\nu^D(t,\tau) \varphi_\nu(\tau),
\end{eqnarray}
where the coefficients $c_\nu^D$ are computed from overlap integrals between the basis functions of the $N$- and $N-1$-electron systems at the respective time steps.  

The integrals in Eq. (\ref{nonad_pw_mo}) involving plane waves $\psi(\textbf{k}_i)$, which correspond to inverse Fourier transforms of the respective Dyson or anion molecular orbitals, can thus be reduced to integrals of the type given in Eq. (\ref{ft_gauss}), e.g.,
\begin{eqnarray}
\langle \psi(\textbf{k}_i)|\psi^D(t,\tau)\rangle=\sum_\nu c_\nu^{D} \langle \textbf{k}_i|\nu\rangle,
\end{eqnarray}
while the overlap integrals $\langle\phi_n|\psi^D\rangle$ between anion MOs and Dyson orbitals reduce to overlaps between Gaussian basis functions.

\subsection{\label{tsh}Quantum-classical surface hopping dynamics}

Having established the equations of motion for the nuclear and electronic degrees of freedom, Eq. (\ref{newton}) and Eqs. (\ref{schroedinger2})/(\ref{schroedinger3}), as well as the necessary energies and couplings, the coupled electron-nuclear dynamics can be described using the surface hopping methodology. Solving Eqs. (\ref{schroedinger2})/(\ref{schroedinger3}) along the nuclear trajectories provides us with the time-dependent electronic state coefficients $c_i(t)$. These are employed in a stochastic process to decide if a switch from the anionic state in which the trajectories are propagated to any of the states of the discretized ionization continuum occurs. Specifically, in every nuclear time step a hopping probability is calculated which depends on the electronic state populations $\rho_{ii}=|c_i|^2$ according to
\begin{eqnarray}
P_{i\rightarrow j}
= 
-\frac{\dot{\rho}_{ii}}{\rho_{ii}} 
\frac{\dot{\rho}_{jj}}{\sum_k \dot{\rho}_{kk}} \Delta t,
\label{probs}
\end{eqnarray}
for $\dot{\rho}_{ii} < 0$ (decrease of initial state population) and $\dot{\rho}_{jj}>0$ (increase of final state population).\cite{lisinetskaya11,domckebook} The sum over $k$ in the denominator extends over all possible final states with a growing population. For all other cases, the hopping probability is set to zero. 

As a result of the hopping procedure, we are provided with the instant of time in which the autoionization takes place, as well as with the specific kinetic energy and k-vector of the generated free electron.

To ensure the energy conservation of the system, a hop is performed only if the total energy of a given trajectory (the anion's electronic energy $E_i^{\textrm{(a)}}$ plus kinetic energy $T^{\textrm{(a)}}$) is at least equal to the final state electronic energy (potential energy $E_j^{\textrm{(n)}}$ of the neutral molecule plus kinetic energy $E_{\textrm{el}}(\textbf{k}_i)$ of the free electron), and the kinetic energy of the neutral molecule, $T^{\textrm{(n)}}$ is rescaled accordingly such that
\begin{eqnarray}
T^{\textrm{(n)}}=E_i^{\textrm{(a)}}+T^{\textrm{(a)}}-E_j^{\textrm{(n)}}-E_{\textrm{el}}(\textbf{k}_i).
\end{eqnarray}

Finally, from the hopping times of the individual trajectories a time-dependent anion population is generated by averaging over the full ensemble of trajectories. 

\subsection{\label{el_ion}Approximate description of adiabatic ionization}
While the main focus of the present work lies on the description of vibration-induced autoionization, which is a nonadiabatic process, the possibility of a purely electronic mechanism without the exchange of energy between the electronic and nuclear degrees of freedom needs to be considered as well. Such mechanism, which we term \textit{adiabatic} in the following, implies that during the course of a trajectory the electron detachment energy may become negative as a result of gradual changes of the nuclear geometry, i.e. the system gets unstable with respect to electron loss. In this situation, one of the system's electrons will form a free wavepacket which will rapidly spread in space, giving rise to a decreasing electron density near the cationic core. In order to obtain an approximate measure of the time scale of this ionization process, we have employed the following procedure for several sample trajectories: For each occurence of a negative VDE in the given trajectory, we take the HOMO of the last step where the electron was still bound ($t_{\mathrm{initial}}$) and consider it as the initial free electron wavepacket. The latter is then propagated freely, and the expectation value of $\hat{r}^2$ (electronic spatial extent) as a function of time is calculated as a measure of the wavepacket spreading as detailed in Appendix \ref{App_C}. To relate this quantity in a simple way to a gradual population loss due to ionization, we consider the following model: The actual wavepacket is replaced by a $1s$-like spherically symmetric electron distribution giving rise to the same value of $\langle\hat{r}^2\rangle$. At $t_{\mathrm{initial}}$, for this distribution the radius of a sphere containing 99\% of the probability is calculated. Subsequently, for each time step the integrated probability within the sphere is computed for the broadening distribution. As a result, a population decay curve is obtained, from which a half-life is determined. The average half-life obtained in these calculations is then employed in the actual simulations of the trajectory ensemble as a time constant to model an exponential population decay due to adiabatic ionization.

\section{\label{sec:Comp}Computational Details}

\begin{table*}
    \caption{\label{tab:table2}
    Comparison of adiabatic electron affinities (AEA), structural parameters and vibrational wavenumbers of vinylidene. The normal modes are denoted as: $\nu_1$: symmetric C-H stretch, $\nu_2$: C-C stretch, $\nu_3$: CH$_2$ scissoring, $\nu_4$: out-of-plane bending, $\nu_5$: antisymmetric C-H stretch, $\nu_6$: CH$_2$ rocking. Normal mode wavenumbers in parentheses are harmonic, the remaining ones anharmonic.\\
    $^\mathrm{a}$ Ref. \onlinecite{stanton99}, 
    $^\mathrm{b}$ Ref. \onlinecite{devineScience}, 
    $^\mathrm{c}$ Ref. \onlinecite{gerardi}, 
    $^\mathrm{d}$ Ref. \onlinecite{isomerization3}
    }
    \begin{ruledtabular}
	\begin{tabular}{ccccccccccc}
		&&&&&\multicolumn{6}{c}{vib. modes / cm$^{-1}$}\\
		&AEA / eV&r$_{\text{CC}}$ / \AA&r$_{\text{CH}}$ / \AA&$\alpha_{\mathrm{HCC}}$&$\nu_1$&$\nu_2$&$\nu_3$&$\nu_4$&$\nu_5$&$\nu_6$\\
	\colrule
		\textbf{$\omega$B97XD} &&&&&&&&&&\\
		6-311++G** & 0.5760 & 1.337 & 1.108 & 123.6 & 2632 & 1485 & 1316 & 768 & 2580 & 869\\
		aug-cc-pVDZ  & 0.5830 & 1.341 & 1.111 & 123.5 & 2622 & 1470 & 1294 & 734 & 2574 & 847\\
		d-aug-cc-pVDZ & 0.5865 & 1.341 & 1.111 & 123.5 & 2621 & 1470 & 1292 & 732 & 2571 & 844\\
		 &  &  &  &  & (2866) & (1531) & (1339) & (775) & (2839) & (874)\\
		d-aug-cc-pVTZ & 0.5434 & 1.332 & 1.105 & 123.6 & 2628 & 1488 & 1309 & 767 & 2571 & 861\\
		\colrule
		\textbf{CAM-B3LYP} &&&&&&&&&&\\
		6-311++G**  & 0.6350 & 1.333 & 1.107 & 123.6 & 2686 & 1496 & 1325 & 780 & 2626 & 875\\
		d-aug-cc-pVDZ  & 0.6463 & 1.341 & 1.112 & 123.5 & 2694 & 1497 & 1304 & 777 & 2647 & 865 \\
		\colrule
		\textbf{LC-$\omega$PBE} &&&&&&&&&&\\
		6-311++G**  & 0.6931 & 1.333 & 1.105 & 123.4 & 2759 & 1525 & 1329 & 797 & 2739 & 883\\
		d-aug-cc-pVDZ  & 0.6992 & 1.341 & 1.110 & 123.4 & 2767 & 1541 & 1312 & 797 & 2774 & 880 \\
	\colrule
		\textbf{CCSD} &&&&&&&&&&\\
		aug-cc-pVDZ  & 0.3911 & 1.364 & 1.117 & 123.5 & (2894) & (1513) & (1341) & (784) & (2877) & (888) \\
		d-aug-cc-pVDZ  & 0.3961 & 1.364 & 1.117 & 123.5 & (2892) & (1512) & (1340) & (777) & (2875) & (885) \\
	\colrule
		\textbf{CCSD(T)} &&&&&&&&&&\\ aug-cc-pVDZ\footnotemark[1] && 1.3668 & 1.1187 & 123.53 & 2667 & 1462 & 1289 & 748 & 2621 & 846\\
	\colrule
		\textbf{Experiment} & 0.4866\footnotemark[2] &&&& 2663\footnotemark[3] & 1485\footnotemark[4] & 1305\footnotemark[4] && 2606\footnotemark[3] &\\
	\end{tabular}
	\end{ruledtabular}
\end{table*}

\begin{figure}[t]
\includegraphics[width=\columnwidth]{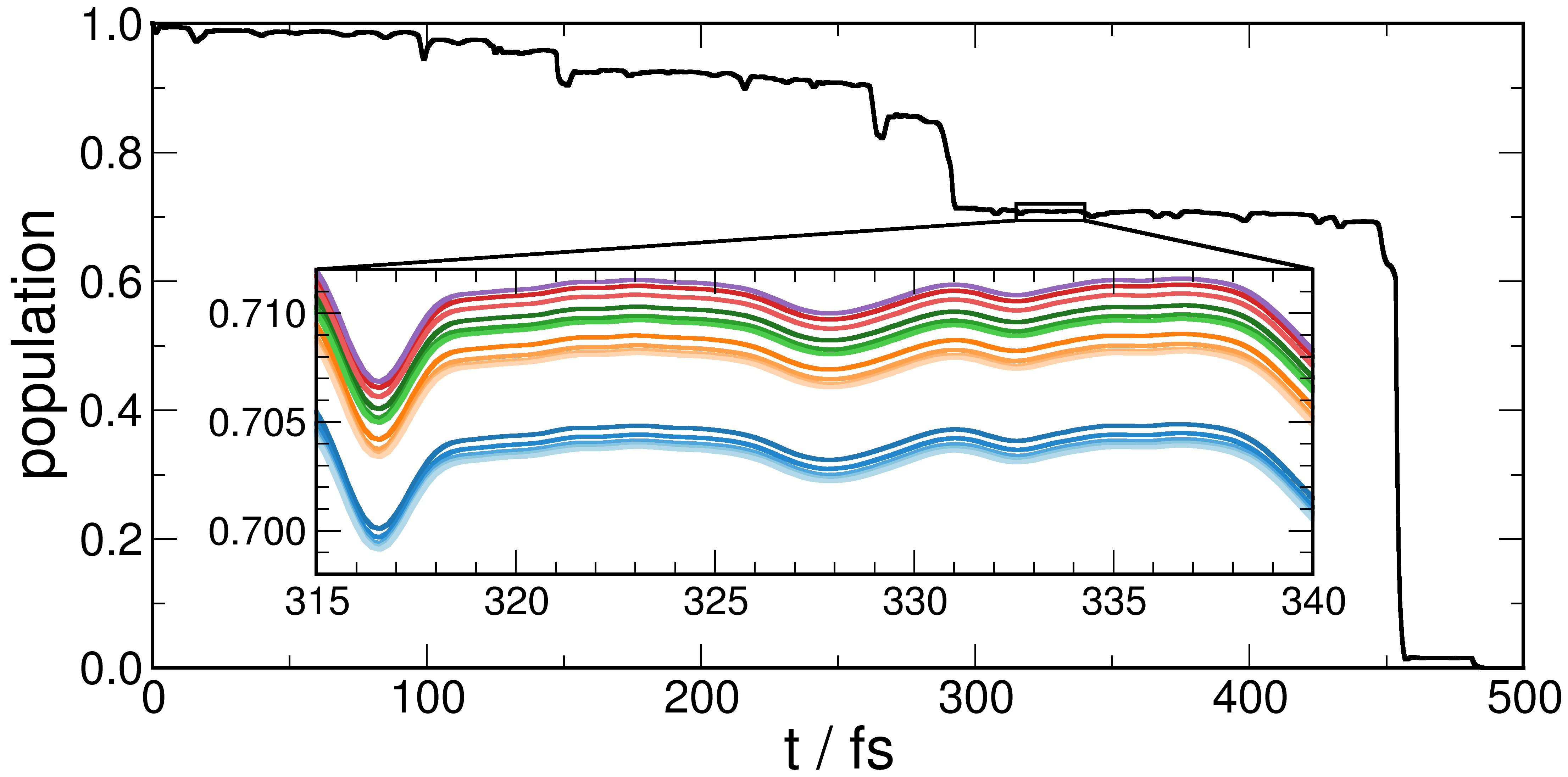}
\caption{\label{fig:comp_discr} Anion populations for a sample trajectory run with different numbers of discretized continuum states according to Table \ref{tab:table1}. The colors represent the discretization scheme as follows: Blue, orange, green, red, and violet correspond to 24, 48, 96, 192, and 384 orientations, while the number of energies in each group increases from darker to lighter colors.}
\end{figure}

The electronic structure of vinylidene was described using density functional theory (DFT). Although the molecule is in principle small enough to afford the use of more accurate ab initio methods, our goal of simulating the dynamics over long time durations in the picosecond regime, as well as the potential applicability of our method to larger molecules, requires the use of a computationally efficient method. In order to serve this purpose in the optimal way, various combinations of the long-range-corrected functionals $\omega$B97XD\cite{wb97xd}, LC-$\omega$PBE\cite{lcwpbe} and CAM-B3LYP\cite{camb3lyp} and the basis sets 6-311++G**\cite{6311ppGss,631pGss3}, (d)aug-cc-pVDZ\cite{augccpvxz1,augccpvxz2,daugccpvxz} and (d)aug-cc-pVTZ\cite{augccpvxz1,augccpvxz2,daugccpvxz} were employed to calculate the geometries, energetics and harmonic normal modes of both the anion and the neutral molecule within the Gaussian 09 program package\cite{g09}. The detailed results are presented in Table \ref{tab:table2} together with data obtained with CCSD\cite{ccsd1,ccsd2}/(d-)aug-cc-pVDZ\cite{augccpvxz1,augccpvxz2,daugccpvxz} and experimental and CCSD(T) data from the literature.\cite{gerardi,isomerization3,stanton99} Inspection of Table \ref{tab:table2} makes clear that among the DFT functionals, $\omega$B97XD provides the best agreement to the experimental and higher-level theoretical data. In an attempt to balance computational cost and the capability of the employed method to properly describe the spatially diffuse electron distribution of the vinylidene anion, we chose to combine $\omega$B97XD with the d-aug-cc-VDZ basis set\cite{daugccpvxz} for use in the trajectory calculations. The initial conditions for all dynamics simulations have been obtained by sampling a quantum phase space distribution. Since the autoionization takes place after vibrational excitation of the molecule, we determined the initial conditions from harmonic normal mode displacements according to the distribution function $P^\nu_\upsilon(Q_\nu,P_\nu)=|\chi^\nu_\upsilon(Q_\nu)|^2|\tilde{\chi}^\nu_\upsilon(P_\nu)|^2$, where $\chi^\nu_\upsilon(Q_\nu)$ and $\tilde{\chi}^\nu_\upsilon(P_\nu)$ are the harmonic oscillator wavefunctions of normal coordinate $\nu$ in position and momentum space, respectively. We set $\upsilon=1$ for selected normal modes according to the experimental findings, and $\upsilon=0$ otherwise. Specifically, we considered the situation where both a single quantum of the C-C stretching ($\nu_2$) and of the antisymmetric C-H stretching mode ($\nu_5$) are excited, corresponding to the most intense autoionization resonance K observed by DeVine $et\ al$\cite{devineScience,devineJPCL}, and  propagated 100 trajectories for a total simulation time of 3 ps using the d-aug-cc-pVDZ basis set.\\
The propagation of the nuclei was performed by numerically solving Newton's equations of motion using the velocity Verlet algorithm\cite{veloverlet} with a time step of 0.2 fs. 
By solving the time-dependent Schrödinger equation in the manifold of the electronic ground state of the vinylidene anion and a large number of discretized continuum states corresponding to the neutral ground state and the detached electron (as detailed below) the electronic degrees of freedom were propagated using Adams's method as implemented in the \texttt{ode} class of Python's \texttt{scipy.integrate} module\cite{scipy} with a time step of $2\cdot 10^{-3}$ fs.
For the description of the continuum states, an evenly spaced grid of kinetic energies between 0.0 and 1.5 eV was employed. For each kinetic energy, the spatial orientations were chosen to evenly cover a spherical surface according to the Fibonacci sphere distribution. The quality of different discretization schemes was assessed by running a sample trajectory with identical initial conditions for various total numbers of kinetic energies and orientations, as summarized in Table \ref{tab:table1}. The outcome in terms of anion populations are shown in Fig. \ref{fig:comp_discr} and make clear that generally, very similar results are obtained for the tested parameters. The largest population difference is less than 1 \% between the settings employing the most vs. the least number of plane waves. Besides the settings with 24 orientations, which stand a bit off (blue curves in Fig. \ref{fig:comp_discr}), the population differences for the other cases are even smaller, around 0.4 \%. As a compromise between a reasonable number of plane waves and computational efficiency, we finally chose a total of 1000 energies and 96 orientations per energy, thus a number of 96000 k-vectors, for the simulation of the complete trajectory ensemble. This corresponds to the middle green curve in Fig. \ref{fig:comp_discr}.

\begin{table}[bt]
\caption{\label{tab:table1}%
Discretization parameters used for a sample trajectory. The boxed "x" indicates the choice of parameters used for the ensemble of trajectories.
}
\begin{ruledtabular}
\begin{tabular}{cccccc}
\textrm{Different} & \multicolumn{5}{c}{Orientations per energy}\\
\textrm{energies}& 24 & 48 & 96 & 192 & 384\\
\colrule
500  & x & x & x & x & x \\
1000 & x & x & \boxed{\text{x}} & x &   \\
2000 & x & x & x &   &   \\
4000 & x & x &   &   &   \\
8000 & x &   &   &   &   \\
\end{tabular}
\end{ruledtabular}
\end{table}

Employing the respective discretization scheme the system of coupled equations (\ref{schroedinger2}) and (\ref{schroedinger3}) was set up, the diabatic and non-adiabatic couplings were evaluated and the state coefficients calculated.
The hopping probabilities were determined in each nuclear time step from the rate of change of the electronic populations according to Eq. (\ref{probs}). 

Since our interest is focused on the course of the ionization process rather than on the fate of the resulting neutral species, we do not propagate the trajectories in the neutral state once a hop has occurred. This in turn allows us to employ a modification of the surface hopping scheme to improve the hopping statistics: Each trajectory is propagated in the anionic state over the full simulation time, and initially, a "trajectory population" of 1000 is assigned to it. In each nuclear time step, hopping is attempted as many times as given by the actual trajectory population, which is then reduced according to the number of successful hops. This procedure is actually equivalent to propagating each set of initial conditions 1000 times. If desired, sequel trajectories in the neutral state could be run nonetheless in order to study the dynamics after the ionization has taken place.

\section{\label{sec:Results}Results and Discussion}
\begin{figure}[t]
\includegraphics[width=\columnwidth]{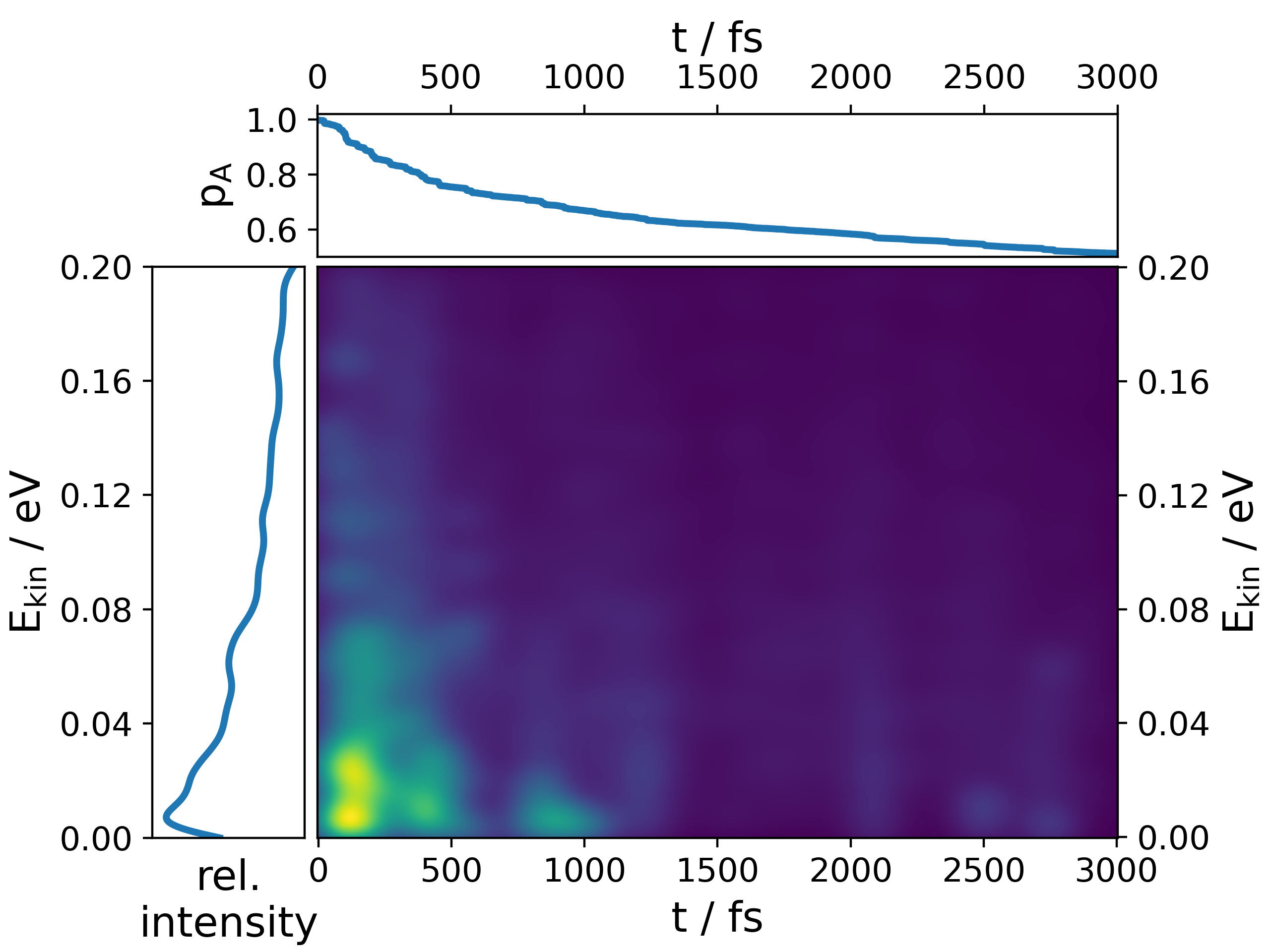}
\caption{\label{fig:2d_daug} Autoionization of vinylidene anions for excitation of the C-C stretching and the antisymmetric C-H stretching vibration ($2_15_1$) using the d-aug-cc-pVDZ basis set. Upper panel: Time-dependent anion population. Left panel: Time-integrated electron kinetic energy distribution. Middle panel: Time-resolved kinetic energy spectrum.}
\end{figure}
Experimentally, autoionization of vinylidene can be induced by infrared excitation of specific normal modes featuring energies above the electron detachment threshold.\cite{gerardi,devineJPCL} For vibrationally cold molecules, this applies notably to combinations of the C-C stretching and the C-H stretching vibrations, which are visible as pronounced resonances in the photodetachment spectrum.\cite{devineJPCL} In our dynamics simulations, we model this situation by sampling initial normal coordinates and momenta from a phase space distribution function accounting for the vibrational excitation as described in the Computational Section.
The coupled electron-nuclear dynamics is then simulated employing Eq. (\ref{newton}) for the nuclei and Eqs. (\ref{schroedinger2}) and (\ref{schroedinger3}) for the electronic degrees of freedom. 

In the following, we consider excitation of the modes $\nu_2$ (C-C stretch) and $\nu_5$ (antisymmetric C-H stretch), which corresponds to the photodetachment peak $K$ in Ref. \onlinecite{devineJPCL} and is henceforth abbreviated as $2_15_1$. The simulation gives rise to autoionization events producing free electrons of specific kinetic energies which can be arranged in a two-dimensional time-resolved kinetic energy spectrum as presented in Fig. \ref{fig:2d_daug}. The plot shows the highest intensity in the time range below 500 fs with energies mostly between 0.0-0.04 eV and a maximum at 0.01 eV. For later times the intensity is weaker and remains maximal around 0.01 eV with almost no intensity above 0.02 eV. Integration over the energies yields the total time-dependent ionization intensity, corresponding to the anion population shown in the upper part of Fig. \ref{fig:2d_daug}, which exhibits a decrease by 50 \% within 3 ps. The time-integrated energy distribution of the ejected electrons presented in the left part of Fig. \ref{fig:2d_daug} exhibits a maximum around 0.01 eV. 

These results can be confronted with the experimental data from Ref. \onlinecite{devineJPCL}, where electrons with a constant kinetic energy of about 115 cm$^{-1}$ (0.0143 eV) were reported to result from excitation of the $2_15_1$ peak in the photodetachment spectrum.
Our kinetic energies are in very good agreement with the experimental data, although more broadly dispersed. This can be expected due to the classical description of the vibrational motion, where no discrete vibrational energy level structure is included in our simulations.
Nonetheless, our approach provides for the first time data on the expected time scales of the autoionization (which is discussed in more detail below) and allows us to analyse the underlying dynamical mechanism. 

\begin{figure}
\includegraphics[width=\columnwidth]{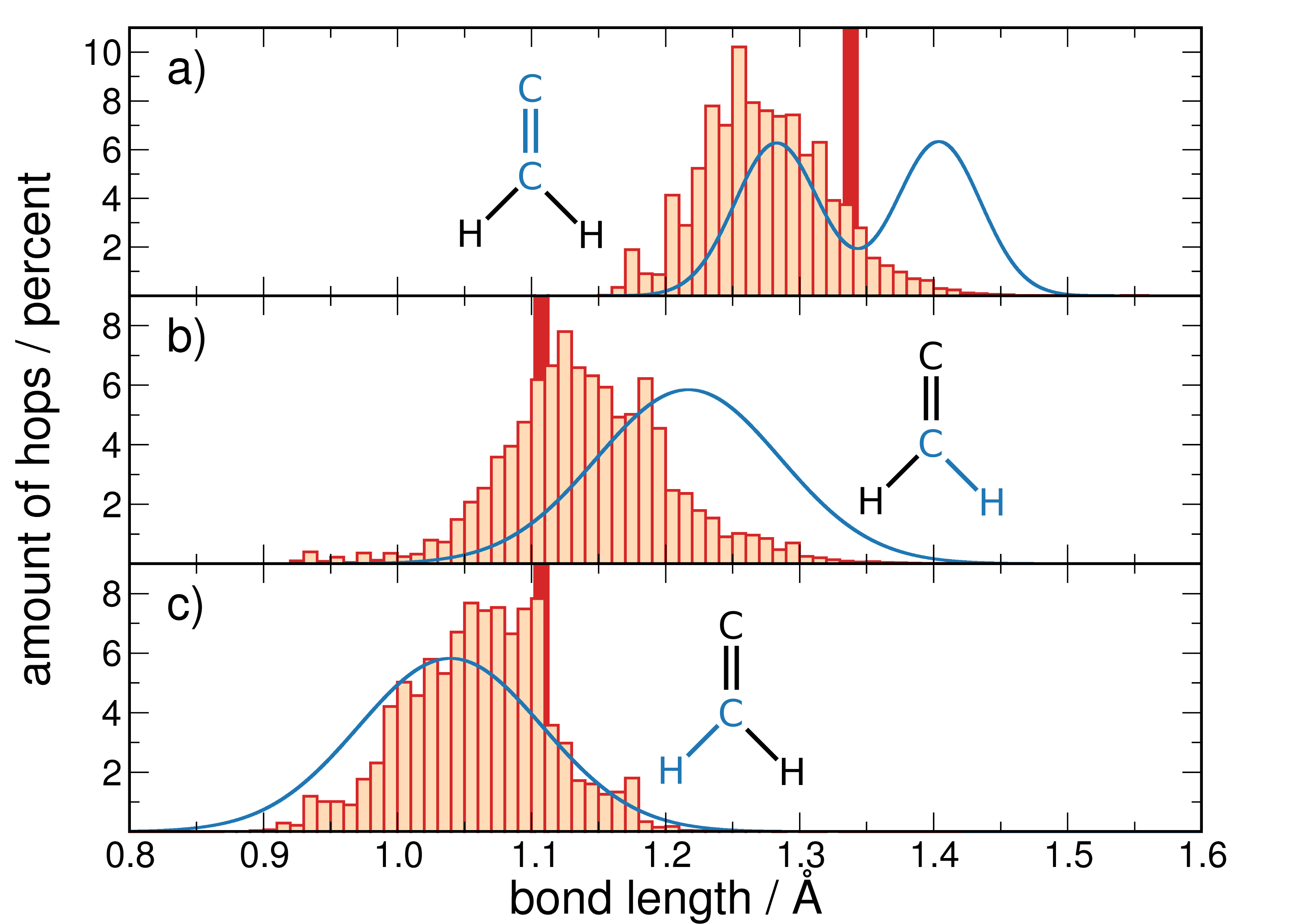}
\caption{\label{fig:hopdist_2m} Distribution of bond lengths at the transition to the ionized state (orange bars) and at $t=0$ (blue curve). The values for the minimum energy structure are given as thick red bars. (a) C-C bond, (b) the longer of the two C-H bonds, (c) the shorter of the two C-H bonds. Notice, that the blue curves do not maximize at the equilibrium values due to the use of vibrationally excited initial conditions.}
\end{figure}

\begin{figure}
\includegraphics[width=\columnwidth]{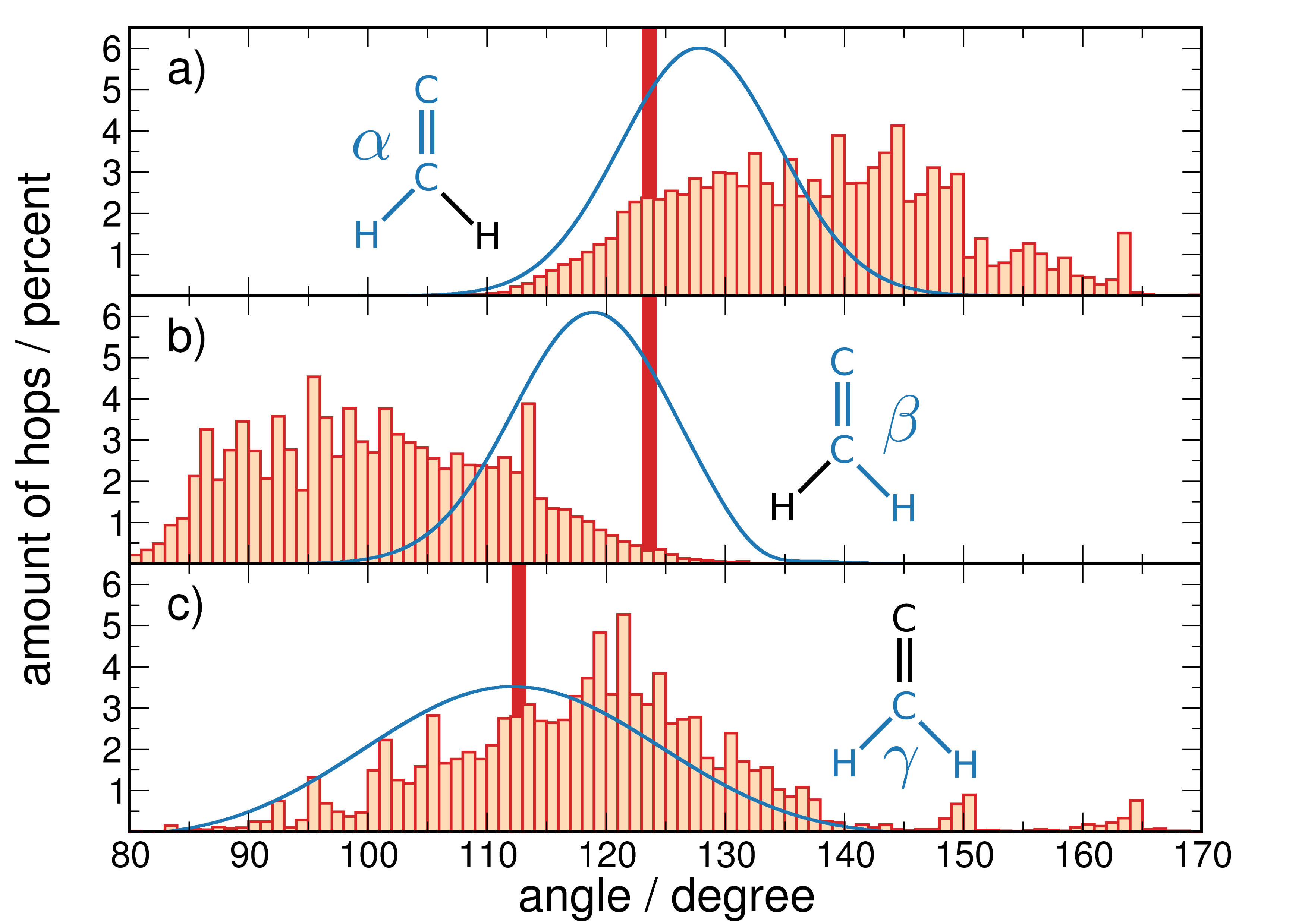}
\caption{\label{fig:hopang_2m} Distribution of bond angles at the transition to the ionized state (orange bars) and at $t=0$ (blue curve). The values for the minimum energy structure are given as thick red bars. (a) larger C-C-H angle, (b) smaller C-C-H angle, (c) H-C-H angle.}
\end{figure}

For this purpose, in the first place we investigate how the molecular structures have changed at the time of the ionization transitions compared to the initial conditions. These changes can be visualized by the distribution of structural parameters (bond lengths, angles) for the whole ensemble of trajectories, as presented in Fig. \ref{fig:hopdist_2m} for the bond lengths and in Fig. \ref{fig:hopang_2m} for several angles. With regard to bond lengths, ionization preferably takes place for shorter values of the C-C bond (Fig. \ref{fig:hopdist_2m}a). This is consistent with the finding that the equilibrium C-C bond length is shorter in neutral vinylidene (anion: 1.34\,{\AA}, neutral: 1.30\,{\AA}), and this situation is even more pronounced in actetylene (1.21\,{\AA}), whose anion is not bound at all (values obtained using DFT, $\omega$B97XD/d-aug-cc-pVDZ). For the C-H bonds there is a tendency avoiding very large and, to a lesser extent, very small values, thus approaching the value for the equilibrium structure (cf. Fig. \ref{fig:hopdist_2m}b and c). For the angles (Fig. \ref{fig:hopang_2m}), there is a clear increase of the larger C-C-H angle ($\alpha$), and a decrease of the smaller one ($\beta$), while the H-C-H angle ($\gamma$) only marginally increases. This can be conceived as the whole CH$_2$ group bending with respect to the C-C bond axis, leading to a T-shaped structure. 

\begin{figure}
\includegraphics[width=\columnwidth]{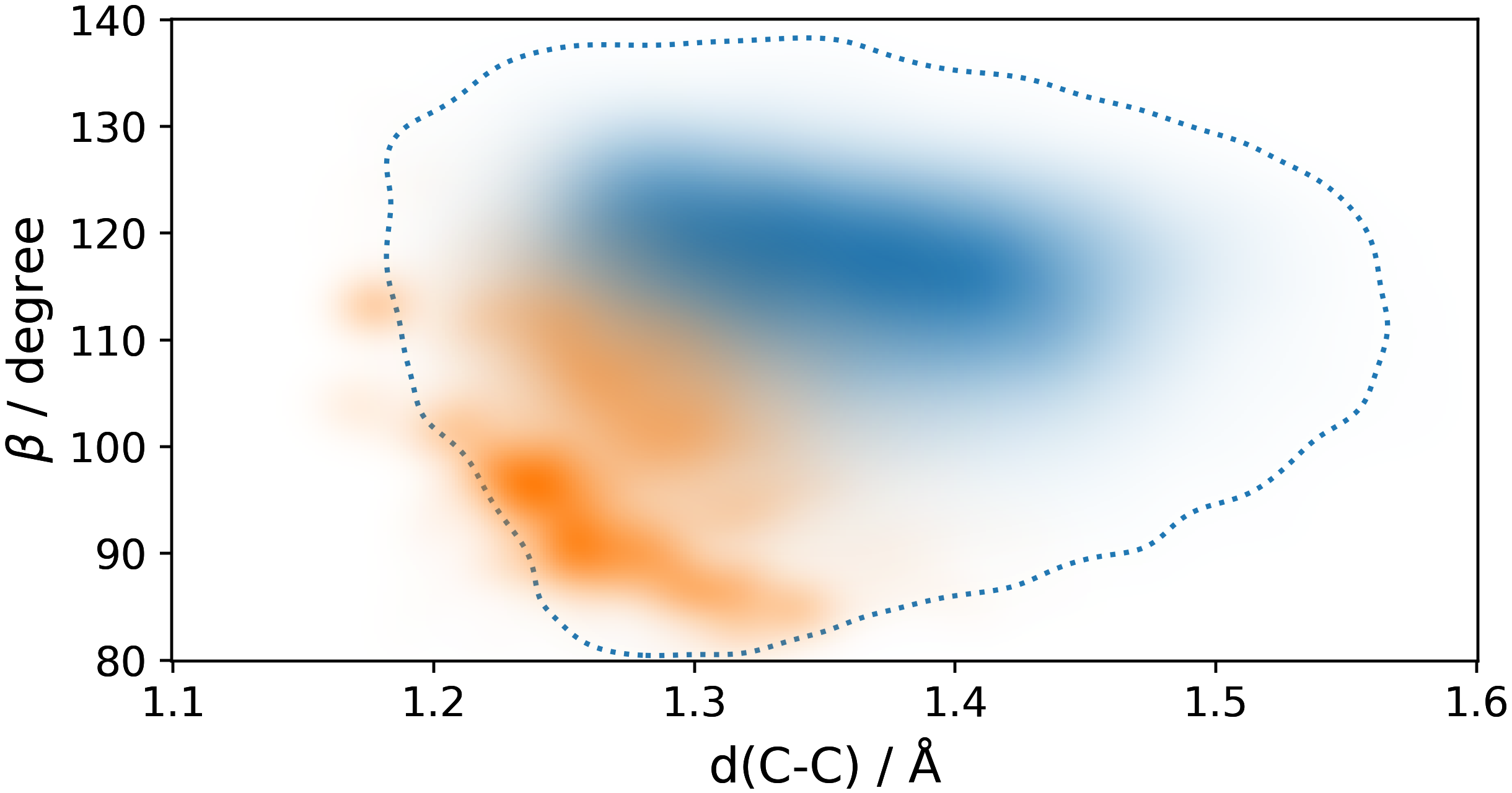}
\caption{\label{fig:betavscc} Correlation of C-C distance and the smaller C-C-H bond angle $\beta$, shown both at the ionization events (orange) and at all instants of time during the dynamics (blue). More intense color indicates more occurrences of the respective bond length/angle combination. The blue dotted curve represents a contour line with 1 \% of the maximum value of the blue distribution.}
\end{figure}

The correlation between the two parameters most distinctly deviating from the initial values, the C-C distance and the C-C-H angle $\beta$, is illustrated more explicitly in Fig. \ref{fig:betavscc}, where the blue area marks the full range of values reached throughout the dynamics simulation while the red area indicates the distribution at the ionization events. The data shows that ionization occurs preferably at geometries with shortened C-C bonds and decreased C-C-H angle. This implies the formation of structures approaching a T-shaped geometry with short C-C bonds, which is known to be the first step in the isomerization from vinylidene to linear acetylene.\cite{carter01} Since anionic acetylene is unstable with respect to electron loss, it is clear that such a process will be accompanied by autoionization. 

\begin{figure}
\includegraphics[width=\columnwidth]{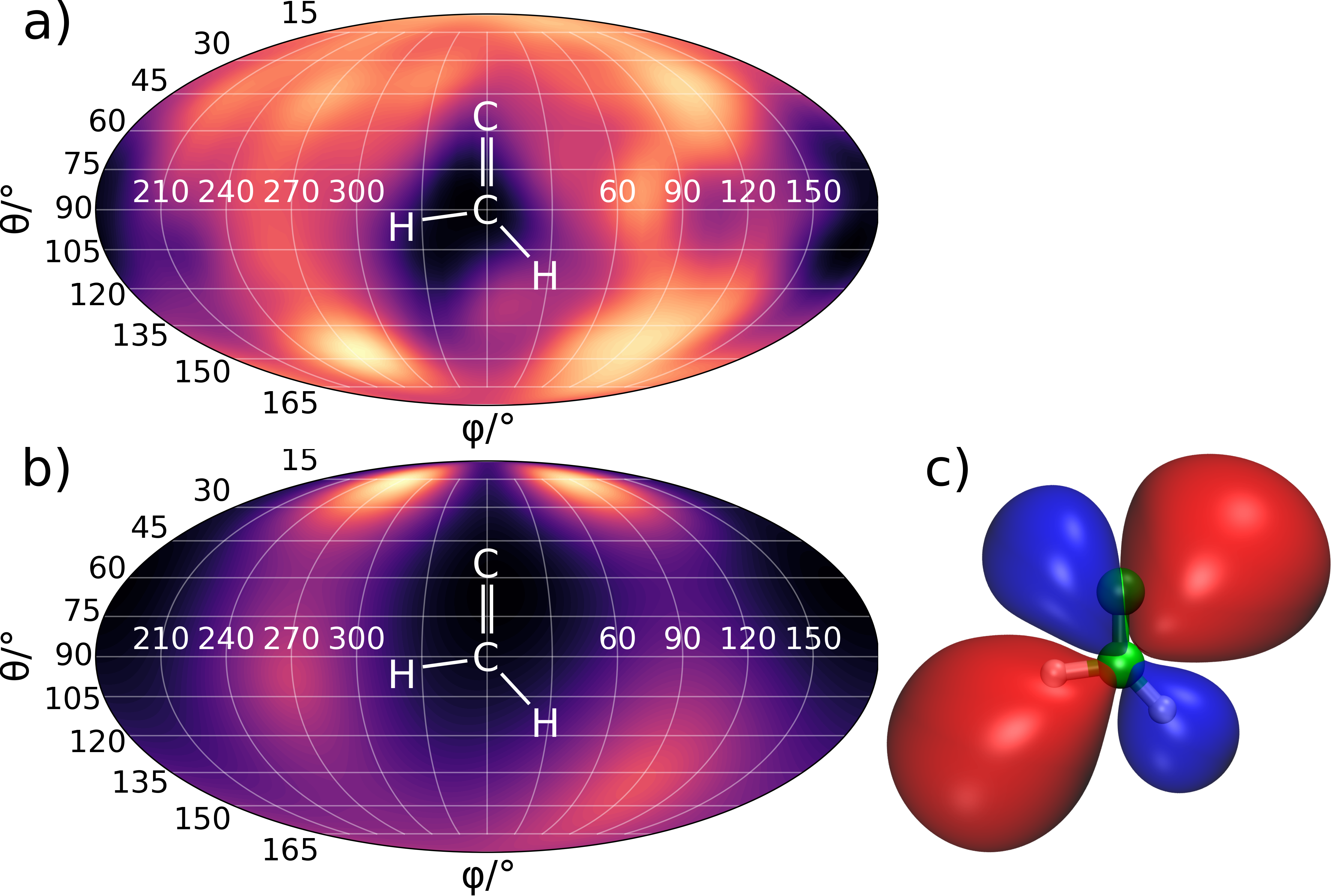}
\caption{\label{fig:contourAngle} (a) Mollweide projection of the angular distribution of ejected electrons summed over all energies. The bright yellow regions indicate large, the dark ones low electron intensity. Overall, the darkest areas feature about 2/3 of the intensity of the brightest ones; (b) Probability density of the Dyson orbital for a mean "ionization geometry" obtained by averaging all structures at which ionization transitions occurred. The electron distribution has been integrated up to 10 {\AA}  from the center of mass of the molecule. (c) Surface plot of the Dyson orbital obtained with a cutoff value of 0.01; the molecular plane represents the yz-plane, to which the direction given by $(\theta,\varphi) = (90^\text{o},0^\text{o})$ is perpendicular.}
\end{figure}

Furthermore, the angular distribution of ejected electrons can be analysed from our simulation data. In Fig. \ref{fig:contourAngle}, this has been accomplished in the form of a Mollweide projection,\cite{map_projections} which is an equal-area, pseudocylindrical map projection of a sphere onto a plane. Each point on the plane corresponds to a direction characterized by the polar and azimuthal angle, $\theta$ and $\varphi$. In the Figure, regions of high electron intensity are marked by bright yellow color, regions of low intensity by dark color. A distinct anisotropy can be observed with electrons primarily ejected at $\varphi$ values around $90^\text{o}$ and $270^\text{o}$, i.e., within the molecular plane. Also for $\theta$, an anisotropy is discernible, with preferred values around $45^\text{o}$ and $150^\text{o}$. This distribution is paralleled by a Dyson orbital calculated for an "average ionization" structure, which has been obtained by averaging all structures at which ionization transitions occurred (see Fig. \ref{fig:contourAngle}b). The corresponding electron distribution is similar to a d-type orbital situated in the molecular plane. Electron ejection preferentially takes place in the directions of the orbital lobes within the molecular plane, while relatively few electrons leave perpendicular to this plane.

\begin{figure}
\includegraphics[width=\columnwidth]{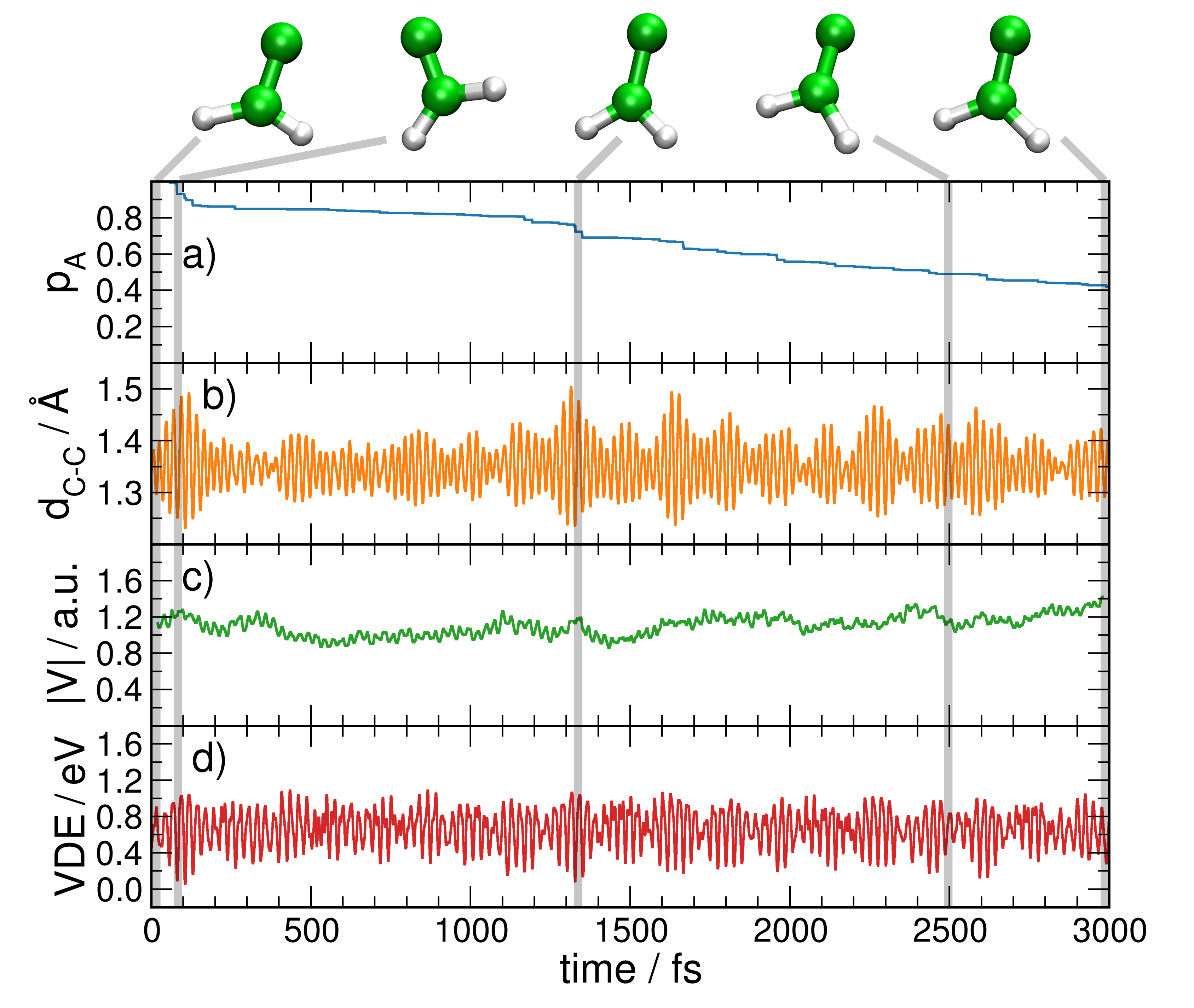}
\caption{\label{fig:2m_traj31} Analysis of a moderately ionizing trajectory. (a) anion population, (b) C-C bond length, (c) average coupling into the ionization continuum, (d) vertical detachment energy (VDE).}
\end{figure}

Having established some structural features promoting autoionization, the question remains at which time scales these are exhibited and how they influence the energetics and couplings governing the ionization efficiency. 
To this end, we analyse in the following the temporal evolution of two sample trajectories in terms of geometric changes and electronic couplings. 
The trajectories have been chosen such that one exhibits a moderate, the other a strong ionization efficiency. 
For the moderately efficient trajectory presented in Fig. \ref{fig:2m_traj31}, the dynamics is characterized by small-amplitude nuclear vibrations without significant structural deformations. 
Ionization events, which are manifest by a dropping anionic population (cf. Fig. \ref{fig:2m_traj31}a), preferably occur at small C-C distances and low VDEs, as indicated by the grey bars at 80 and 1340 fs. 
In these particular regions, the coupling (which is presented as a running average for better comprehensibility) between the bound and ionized states exhibits broad maxima, exceeding the average coupling present at other times. One can also see an overall increased coupling strength with larger times, leading to a gradual increase in population loss for this trajectory.
The final anionic population reaches about 40 \%. 

\begin{figure}
\includegraphics[width=\columnwidth]{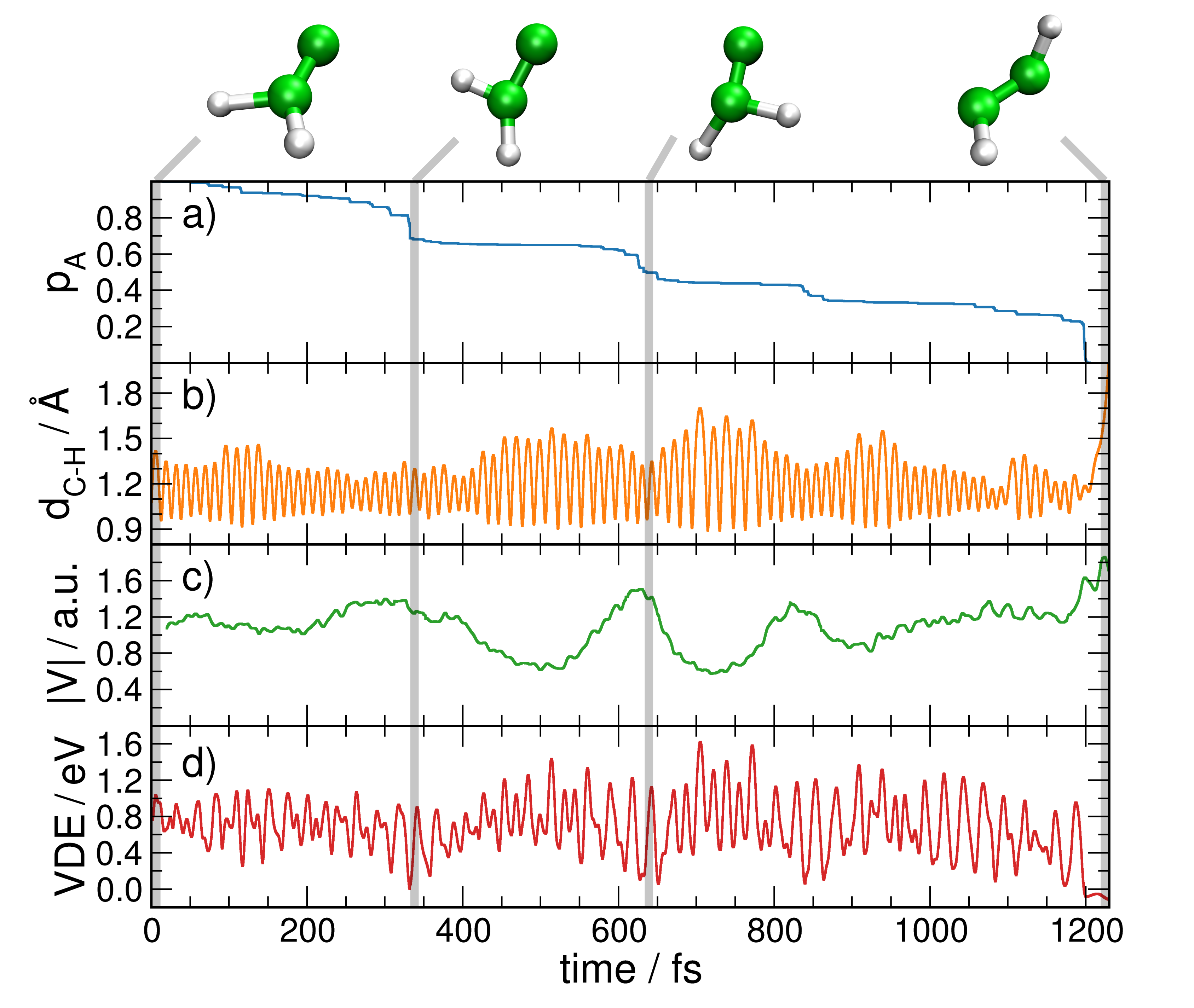}
\caption{\label{fig:2m_traj54} Analysis of a strongly ionizing trajectory. (a) anion population, (b) C-H bond length, (c) average coupling into the ionization continuum, (d) vertical detachment energy (VDE).}
\end{figure}

\begin{figure}
\includegraphics[width=\columnwidth]{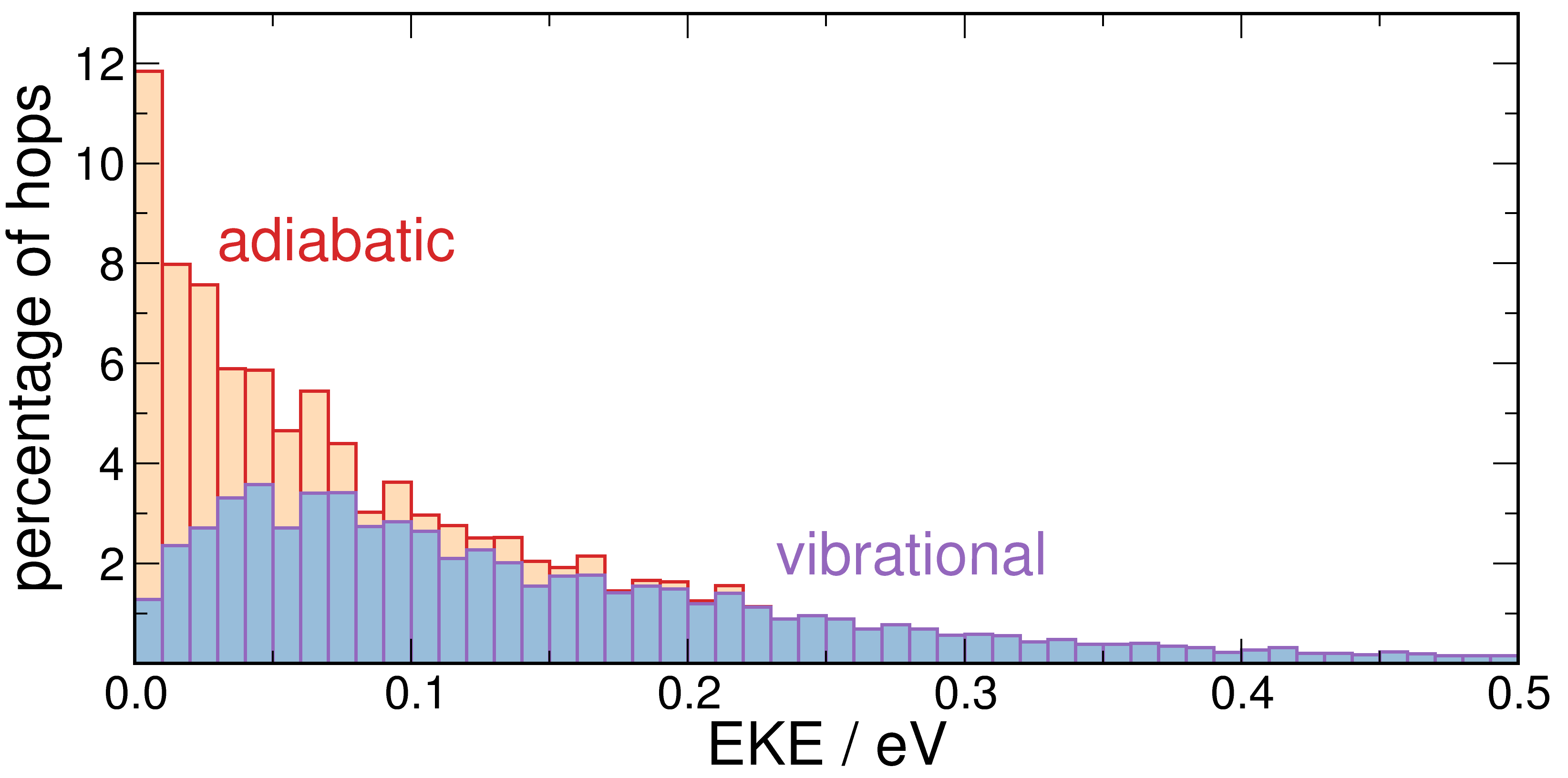}
\caption{\label{fig:ekin_el_nonad} Stacked column chart showing the kinetic energy distribution in terms of hopping events pertaining to the vibrational (positive VDE, blue bars) and adiabatic (negative VDE, orange bars) ionization mechanisms.}
\end{figure}

For the strongly ionizing trajectory illustrated in Fig. \ref{fig:2m_traj54}, the situation is different insofar as large-amplitude nuclear motion takes place. Specifically, the C-H bonds exhibit intense vibrations from 300 fs onwards, until ultimately at 1200 fs a geometry is adopted in which one of the hydrogens migrates from one carbon atom to the other, giving rise to an acetylene-like structure. Since anionic acetylene is electronically unstable, this means that the system should now be actually composed of a neutral acetylene molecule and a free electron. Ionization is achieved efficiently by an "adiabatic" electronic mechanism in this case, i.e. the continuous change in nuclear configuration gradually lowers the VDE until a negative value is reached. The use of a doubly-augmented Gaussian basis set allows for an approximate modelling of this process, although the ultimately the localized nature of the basis set prevents a full description of an electron moving away from the molecule. Instead, an artificial rebound of the electron would be observed for sufficiently long simulation times. What should happen in reality, though, is a fast dispersion of the unbound electron wavefunction, leaving behind neutral acetylene. To grasp an approximate time scale of this ionization mechanism, we have modelled the wavepacket dispersion according to the procedure described in Section \ref{el_ion}, which results in a very fast decay of the anionic population once a negative VDE occurs. Having reached zero population, the respective trajectories are aborted (e.g., at 1200 fs for the trajectory shown in Fig. \ref{fig:2m_traj54}. It should be noted in these cases two different autoionization mechanisms are observed: On the one hand the nonadiabatic vibration-induced autoionization, which is induced by the electronic couplings discussed in Section \ref{nonad} and is accompanied by energy redistribution between the electronic and nuclear degrees of freedom, and on the other hand the aforementioned purely electronic adiabatic mechanism which is active as soon as molecular geometries with a negative VDE are reached (at, e.g., around 330 fs, 630 fs and 1200 fs in Fig. \ref{fig:2m_traj54}). In this situation, the electron configuration itself becomes unstable, and the excess electron can move away without the need of energy gain from the nuclear system. This mechanism occurs in 25 of 100 trajectories and is responsible for 35 \% of all ionization events in the present simulation. The time scale of free electron dispersion is around 1 fs on average. The decomposition of the kinetic energy spectrum with respect to the two mechanisms is illustrated in Fig. \ref{fig:ekin_el_nonad}, showing that the adiabatic mechanism preferably results in low energy electrons.

\begin{figure}
\includegraphics[width=\columnwidth]{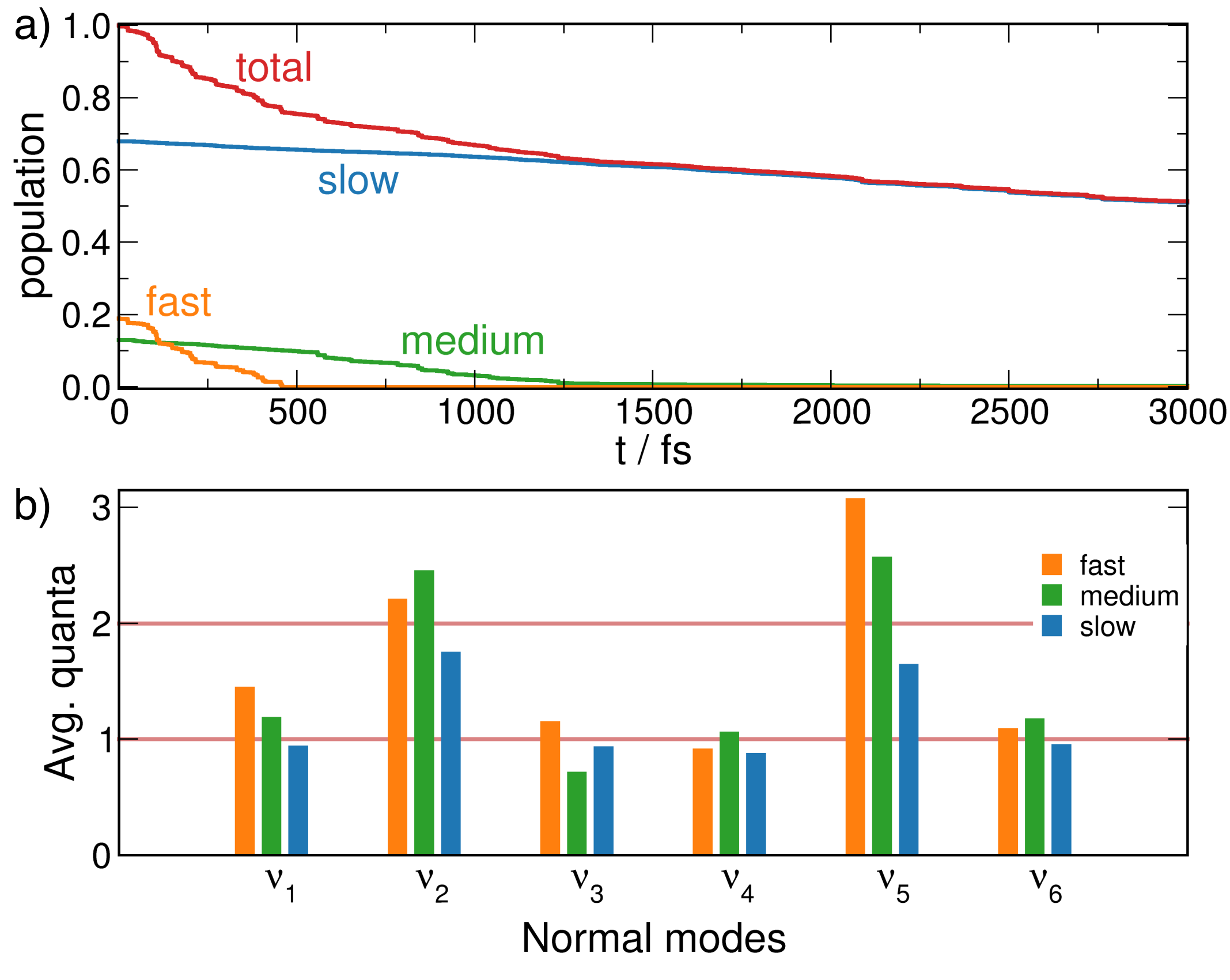}
\caption{\label{fig:pop_group_modes} (a) Decomposition of the total anion population (red) according to the speed of ionization of the underlying trajectories. Fast trajectories (orange) lose 50\% of their population within 500 fs, medium ones (green) within 1500 fs and slow ones (blue) need more than 1500 fs. (b) Group-averaged energy distribution of the initial conditions in terms of normal mode vibrational quanta.}
\end{figure}

The presence of two autoionization mechanisms together with the fact that individual trajectories may ionize on different time scales necessitates a more comprehensive analysis of the temporal characteristics of the ionization process. To this end, the population curve from Fig. \ref{fig:2d_daug} is replotted in Fig. \ref{fig:pop_group_modes}a (red curve). Inspection of this curve already hints at different underlying time scales, as within the first 500 fs the it decreases more steeply than afterwards. 

This finding can be made more quantitative by arranging the trajectories into groups according to the speed of the ionization process. Specifically, we define three groups by asking when the individual trajectory population has decreased below 50 \%: For the fast group, this occurs within the first 500 fs of the dynamics, for the medium group between 500 and 1500 fs, and for the slow group at times beyond 1500 fs. The resulting decomposition of the anionic populations is presented in Fig. \ref{fig:pop_group_modes}a, proving that the initial drop of the total population (red curve) within the first 500 fs is indeed due to only a small fraction of trajectories (orange curve). Besides another part that ionizes in an intermediate time range (green curve), the largest subgroup consists of trajectories ionizing only slowly within the simulation time (blue curve). 

With regard to adiabatic ionization, we find that it is exhibited mostly by trajectories of the fast group (64\%), while the medium and slow groups only amount 24 and 12 \%, respectively. Within the groups, the mechanism occurs in a large majority of the fast trajectories (84\%), while it is much rarer in the medium (46\%) and slow (4\%) groups.

To explain the different behavior of the three groups, we analysed the initial energy distribution of the trajectories in terms of averaged harmonic normal mode quantum numbers as shown in Fig. \ref{fig:pop_group_modes}b. The general appearance of the plotted average quanta per normal mode and trajectory group reflects the fact that the employed initial conditions feature vibrational excitation (i.e. 2 quanta) in modes 2 and 5, while for the others the vibrational ground state was populated (1 quantum). Overall, for the two excited modes 2 and 5 as well as for mode 1, the slow group clearly exhibits the lowest average quantum numbers. For mode 5, the differences are most pronounced, with the fast group being excited on average by about 3 quanta, while for the slow group an average quantum number well below two is observed. However, for the other excited mode (2), differences between the groups are smaller, and the highest average quantum number is observed for the "medium" group, although both the medium and fast groups exhibit average quantum numbers above two. Generally, higher initial quantum numbers correlate with faster ionization, and for mode 2 (C-C bond stretching), this can be directly linked to the fact that for elongated C-C bonds the VDE is reduced. Even stronger reduction of the VDE would be expected for motion towards T-shaped or acetylene-like structures, as promoted by the CH$_2$ rocking mode (mode 6). This mode is not directly excited at the beginning, but seems to be sufficiently strongly coupled to modes 5 and 1 (antisymmetric and symmetric C-H stretch) such that excitation of the latter also contributes to faster ionization time scales.

\section{\label{sec:Concl}Conclusion}
We have presented a generally applicable method for the simulation of vibration-induced autoionization dynamics in molecules. 
Our approach is based on the mixed quantum-classical surface hopping scheme where the nuclei are propagated classically while the electronic degrees of freedom are treated quantum mechanically. 
The electronic states considered include 
(i) the bound states of the molecular anion, which are described using quantum chemical methods employing sufficiently diffuse basis sets, and 
(ii) the ionized system composed of the neutral molecular core (also treated by standard quantum chemical methods) and the free electron, which is approximated by orthogonalized plane waves. 
The ionization continuum is discretized and represented by a large set of individual discrete box-normalized states. 
The electronic couplings necessary to describe the transitions between the bound and ionized molecular states consist of two contributions: 
(i) the nonadiabatic couplings due to the change of the anion and neutral electronic wavefunctions as a function of the nuclear coordinates, and
(ii) a diabatic coupling between anionic and neutral states which results from the fact that the ionized-state wavefunction is not obtained self-consistently, but is constructed from individually computed neutral molecular and free electron wavefunctions.
In addition to the ionization mechanism mediated by these couplings, we also include in our treatment a purely electronic effect which occurs when due to the nuclear motion the anionic wavefunction "adiabatically" becomes unbound without direct coupling to the nuclear degrees of freedom.

We have illustrated our approach by simulating the autoionization dynamics of the vinylidene anion following vibrational excitation. Our results provide for the first time an estimate for the time scale of this process which has been previously studied experimentally\cite{devineJPCL} and allow us to link the ionization efficiency to specific geometrical deformations of the molecules as well as to the choice of initial conditions in terms of vibrational excitation.
Our methodology can be straightforwardly applied to more complex molecules, providing a means to assess the autoionization dynamics for cases well beyond the reach of full quantum wavepacket based simulations. 

\begin{acknowledgments}
We wish to acknowledge financial support by the Deutsche Forschungsgemeinschaft in the frame of the Research Training Group GRK 2112.
\end{acknowledgments}

\newpage
\appendix

\section{Derivation of diabatic coupling}\label{App_A}
In the following, the diabatic coupling elements between a bound anion ground state, $| \Phi_0\rangle$, and a singly-ionized continuum state, $|\Phi_i(\textbf{k}_i)\rangle$, which are both approximated by a single Slater determinant will be derived. The two determinants read:
\begin{eqnarray}
| \Phi_0\rangle&=&\frac{1}{\sqrt{N!}}\begin{vmatrix}
\phi_1(1) & ... & \phi_N(1)\\
... & ... & ...\\
\phi_1(N)&...&\phi_N(N)
\end{vmatrix}\\
|\Phi_i(\textbf{k}_i)\rangle&=&\frac{1}{\sqrt{N!}}\begin{vmatrix}
\tilde{\psi}(\textbf{k}_i,1)&\chi_1(1)&...&\chi_{N-1}(1)\\
...&... & ... & ...\\
\tilde{\psi}(\textbf{k}_i,N)&\chi_1(N)&...&\chi_{N-1}(N)
\end{vmatrix},\nonumber \\
&&
\end{eqnarray}
where $\phi_i$ denotes anion MOs, $\chi_i$ neutral MOs, and $\tilde{\psi}(\textbf{k}_i)$ is the orthogonalized plane wave describing the free electron. For later convenience, we also define the overlap matrix between the two sets of orbitals:
\begin{eqnarray}
\mathbf{S}=\begin{pmatrix}
\langle \tilde{\psi}|\phi_1\rangle & ... & \langle \tilde{\psi}|\phi_N\rangle\\
\langle \chi_1|\phi_1\rangle & ... & \langle \chi_1|\phi_N\rangle\\
... & ... & ...\\
\langle\chi_{N-1}|\phi_1\rangle&...&\langle\chi_{N-1}|\phi_N\rangle
\end{pmatrix}.
\end{eqnarray}
The diabatic coupling can then be written as
\begin{equation}
\begin{split}
V^{\mathrm{dia}}_{i0}(\textbf{k}_i)&=\langle \Phi_i(\textbf{k}_i)|\hat{H}| \Phi_0\rangle\\
&=\sum_{a=1}^N \langle \Phi_i(\textbf{k}_i)|\hat{h}(a)| \Phi_0\rangle+\sum_{a=1}^N\sum_{b=a}^N \langle \Phi_i(\textbf{k}_i)|\hat{v}(a,b)| \Phi_0\rangle\\
&\equiv V^{\mathrm{dia,1}}+V^{\mathrm{dia,2}},
\end{split}\label{diab}
\end{equation}
where $\hat{h}(a)=-\frac{1}{2}\nabla_a^2+v_\mathrm{ne}(a)+v_\mathrm{nn}(a)$ denotes the one-electron part of the Hamiltonian pertaining to electron $a$ and comprises the kinetic energy as well as the potential energies of the electron-nuclear ($v_{\mathrm{ne}}$) and the internuclear interactions ($v_{\mathrm{nn}}$). The potential energy operator for the interaction between electrons $a$ and $b$ is denoted as $\hat{v}(a,b)$.  In the last row of. Eq. (\ref{diab}) the coupling has been formally decomposed into one- and two-electron contributions. For the one-electron part, it can be shown that the matrix element reduces to a form only involving the plane wave $\tilde{\psi}$ and the Dyson orbital $\psi^D=\sqrt{N}\langle \Phi^{N-1}_0|\Phi^{N}_0 \rangle$ for the ionization transition: 
\begin{eqnarray}
V^{\mathrm{dia,1}}&=&\langle \tilde{\psi}|\hat{h}(1)|\sum_p(-1)^{p+1} \mathrm{det}\,\mathbf{S}_{i,p}|\phi_p\rangle\\
&=&\langle \tilde{\psi}|\hat{h}|\psi^D\rangle.
\end{eqnarray}
Before turning to the two-electron part, for brevity we introduce a short-hand notation for the electron-electron interaction integrals,
\begin{eqnarray}
\langle \tilde{\psi}(\mathbf{k}_i)(1)\chi_n(2)|\hat{v}(1,2)|\phi_p(1)\phi_q(2)\rangle\equiv \langle \tilde{\mathbf{k}}_in|pq\rangle, 
\end{eqnarray}
and further, for their antisymmetrized versions,
\begin{eqnarray}
\langle \tilde{\mathbf{k}}_in|pq\rangle-\langle \tilde{\mathbf{k}}_in|qp\rangle \equiv \langle \tilde{\mathbf{k}}_in||pq\rangle.
\end{eqnarray}
The two-electron part can be reduced to integrals involving up to four different MOs:
\begin{eqnarray}
V^{\mathrm{dia,2}}&=&\sum_n \sum_{q,p<q}\langle \tilde{\mathbf{k}}_in||pq\rangle \cdot (-1)^{s} \mathrm{det}\,\mathbf{S}_{in,pq},
\end{eqnarray}
with $s=n+p+q-1$ and $\mathrm{det}\,\mathbf{S}_{in,pq}$ as the minor determinant of matrix $\mathbf{S}$ where the rows $i$ and $n$ as well as the columns $p$ and $q$ have been deleted. In order to further simplify this expression, we expand the neutral MOs $\chi_n$ with respect to the anionic ones, $\phi_u$, which leads to
\begin{eqnarray}
V^{\mathrm{dia,2}}&=&\sum_n^{\mathrm{occ}} \sum_u^{\mathrm{all}} \langle \chi_n|\phi_u\rangle \sum_{q,p<q}^{\mathrm{occ}} \langle \tilde{\mathbf{k}}_iu||pq\rangle  (-1)^{s} \mathrm{det}\,\mathbf{S}_{in,pq}\nonumber\\
&&\\
&=&\sum_u^{\mathrm{all}} \sum_{q,p<q}^{\mathrm{occ}} \langle \tilde{\mathbf{k}}_iu||pq\rangle \sum_n^{\mathrm{occ}} S_{nu}(-1)^{s} \mathrm{det}\,\mathbf{S}_{in,pq}\nonumber\\\label{vdia2_det}
\end{eqnarray}

If $\phi_u$ is an occupied orbital, the last sum in Eq. (\ref{vdia2_det}) can be further simplified:
\begin{eqnarray}
&&\sum_n^{\mathrm{occ}} S_{nu}\cdot (-1)^{n+p+q-1} \mathrm{det}\,\mathbf{S}_{in,pq}\\
&=&\begin{cases}
(-1)^{q}\,\mathrm{det}\,\mathbf{S}_{i,q} & u=p\\
(-1)^{p-1}\,\mathrm{det}\,\mathbf{S}_{i,p} & u=q\\
0 & u\neq p,q
\end{cases}
\end{eqnarray}

With this, the two-electron part of the coupling can be decomposed into an occupied and a virtual part as 

\begin{eqnarray}
V^{\mathrm{dia,2}}_{\mathrm{occ}}&=&\sum_{q,p<q}^{\mathrm{occ}} \left (\langle \tilde{\mathbf{k}}_ip||pq\rangle  (-1)^{q}\,\mathrm{det}\,\mathbf{S}_{i,q}\nonumber\right.\\
&&\left.+\langle \tilde{\mathbf{k}}_iq||pq\rangle  (-1)^{p-1}\,\mathrm{det}\,\mathbf{S}_{i,p}\right)\\
V^{\mathrm{dia,2}}_{\mathrm{virt}}&=& \sum_u^{\mathrm{virt}} \sum_{q,p<q}^{\mathrm{occ}} \langle \tilde{\mathbf{k}}_iu||pq\rangle \sum_n^{\mathrm{occ}} S_{nu}\cdot(-1)^{s} \mathrm{det}\,\mathbf{S}_{in,pq}\nonumber\\ 
\end{eqnarray}

The occupied part can be reformulated by setting $\sum_{q,p<q}\rightarrow\frac{1}{2}\sum_{pq}$ and interchanging the indices $p$ and $q$ in the first summand, giving rise to

\begin{eqnarray}
V^{\mathrm{dia,2}}_{\mathrm{occ}}&=&\frac{1}{2}\sum_{pq}^{\mathrm{occ}} \left (\langle \tilde{\mathbf{k}}_iq||qp\rangle  (-1)^{p}\,\mathrm{det}\,\mathbf{S}_{i,p}\nonumber\right.\\
&&\left.+\langle \tilde{\mathbf{k}}_iq||pq\rangle  (-1)^{p-1}\,\mathrm{det}\,\mathbf{S}_{i,p}\right)\\
&=&\sum_{pq}^{\mathrm{occ}} \langle \tilde{\mathbf{k}}_iq||pq\rangle  (-1)^{p+1}\,\mathrm{det}\,\mathbf{S}_{i,p}\\
&=&\sum_{p}^{\mathrm{occ}}\langle\tilde{\mathbf{k}}_i|\sum_q^{\mathrm{occ}}(\hat{J}_q-\hat{K}_q)|\phi_p\rangle(-1)^{p+1}\,\mathrm{det}\,\mathbf{S}_{i,p}\nonumber\\
\end{eqnarray}

Together with the one-electron part we get

\begin{eqnarray}
&&V^{\mathrm{dia,1}}+V^{\mathrm{dia,2}}_{\mathrm{occ}}\nonumber\\
&=&\sum_p^{\mathrm{occ}}\langle\tilde{\mathbf{k}}_i|\hat{h}+\sum_q^{\mathrm{occ}}(\hat{J}_q-\hat{K}_q)|\phi_p\rangle(-1)^{p+1}\,\mathrm{det}\,\mathbf{S}_{i,p}\nonumber\\
&&\\
&=&\sum_p^{\mathrm{occ}}\langle\tilde{\mathbf{k}}_i|\hat{f}|\phi_p\rangle(-1)^{p+1}\,\mathrm{det}\,\mathbf{S}_{i,p},
\end{eqnarray}
where $\hat{f}$ denotes the Fock operator. If the $\phi_p$ are Hartree-Fock orbitals, $\hat{f}|\phi_p\rangle=\varepsilon_p|\phi_p\rangle$ and the complete expression $V^{\mathrm{dia,1}}+V^{\mathrm{dia,2}}_{\mathrm{occ}}$ becomes identically zero due to the orthogonality between the free electron wavefunction $\tilde{\psi}(\mathbf{k}_i)$ and the MOs $\phi_p$. In our case, by contrast, the $\phi_p$ are Kohn-Sham orbitals. However, in our simulations these are found to be very close to the Hartree-Fock orbitals obtained using the same basis set, thus we still set as an approximation 
\begin{eqnarray}
V^{\mathrm{dia,1}}+V^{\mathrm{dia,2}}_{\mathrm{occ}}&\approx{}&0
\end{eqnarray}

Therefore, the expression for the diabatic coupling in the MO basis, as also given in Eq. (\ref{diabatic_twoel}) of the main text, reads
\begin{eqnarray}
V^{\mathrm{dia}}_{i0}(\textbf{k}_i)&=&\sum_u^{\mathrm{virt}} \sum_{q,p<q}^{\mathrm{occ}} \langle \tilde{\mathbf{k}}_iu||pq\rangle\sum_n^{\mathrm{occ}} S_{nu}(-1)^{s} \mathrm{det}\,\mathbf{S}_{in,pq}.\nonumber\\ 
\end{eqnarray}
If the $N$-electron system is a spin doublet anion and the electron ejected in the ionization process has $\alpha$ spin, the $p$ sum in the above formula becomes restricted to $\alpha$ orbitals only. In addition if $q$ represents a $\beta$ orbital all exchange terms in the electron-electron repulsion integrals are also zero.
Inserting subsequently the expression for the orthogonalized plane wave, 
\begin{eqnarray}
\langle \tilde{\mathbf{k}}_i|=\langle \mathbf{k}_i|-\sum_r^{\mathrm{occ},\alpha}\langle \mathbf{k}_i|r\rangle\langle r|,
\end{eqnarray}
with $r$ denoting occupied anion orbitals, yields
\begin{alignat}{1}
V^{\mathrm{dia}}_{i0}(\textbf{k}_i)
=
\sum_{p}^{\mathrm{occ},\alpha}
    \Bigg[&
        \sum_n^{\mathrm{occ},\alpha}
            \sum_{q>p}^{\mathrm{occ},\alpha}
                (-1)^{s} \mathrm{det}\ \mathbf{S}_{in,pq}
                \text{K}\nonumber\\
        +&
        \sum_{\bar{n}}^{\mathrm{occ},\beta}
            \sum_{\bar{q}}^{\mathrm{occ},\beta}
                (-1)^{\bar{s}} \mathrm{det}\ \mathbf{S}_{i\bar{n},p\bar{q}}
                \bar{\text{K}}
    \Bigg]\label{vdia_mo_pw}
\end{alignat}
with
\begin{alignat}{1}
\text{K}
=&
\sum_u^{\mathrm{virt},\alpha}
    S_{nu}
    \Big(
        \langle
            \mathbf{k}_iu||pq
        \rangle
        -
        \sum_r^{\mathrm{occ},\alpha}
            \langle 
                \mathbf{k}_i|r
            \rangle
            \langle 
                ru||pq
            \rangle
    \Big)\\
\bar{\text{K}}
=&
\sum_{\bar{u}}^{\mathrm{virt},\beta}
    S_{\bar{n}\bar{u}}
    \Big(
        \langle
            \mathbf{k}_i\bar{u}|p\bar{q}
        \rangle
        -
        \sum_r^{\mathrm{occ},\alpha}
            \langle 
                \mathbf{k}_i|r
            \rangle
            \langle 
                r\bar{u}|p\bar{q}
            \rangle
    \Big)
\end{alignat}
where $\langle \mathbf{k}_i|$ is the pure plane wave of wavevector $\mathbf{k}_i$.

A final technical simplification can be achieved by noticing that there are usually many more virtual than occupied orbitals. Therefore, we use the expansion of the neutral MOs $\langle n|$ with respect to the anionic ones, $\langle u|$, to set
\begin{equation}
\sum_u^{\mathrm{virt}} S_{nu}\langle u|=\langle n|-\sum_u^{\mathrm{occ}} S_{nu}\langle u|,
\end{equation}
thus avoiding the summation over all virtual anion MOs: 
\begin{alignat}{1}
\text{K}
=&
\langle
    \mathbf{k}_in||pq
\rangle
-
\sum_r^{\mathrm{occ},\alpha}
    \langle 
        \mathbf{k}_i|r
    \rangle
    \langle 
        rn||pq
    \rangle
-\nonumber\\
&\sum_u^{\mathrm{occ},\alpha}
    S_{nu}
    \Big(
        \langle
            \mathbf{k}_iu||pq
        \rangle
        -
        \sum_r^{\mathrm{occ},\alpha}
            \langle 
                \mathbf{k}_i|r
            \rangle
            \langle 
                ru||pq
            \rangle
    \Big)\\
\bar{\text{K}}
=&
\langle
    \mathbf{k}_i\bar{n}|p\bar{q}
\rangle
-
\sum_r^{\mathrm{occ},\alpha}
    \langle 
        \mathbf{k}_i|r
    \rangle
    \langle 
        r\bar{n}|p\bar{q}
    \rangle
-\nonumber\\
&\sum_{\bar{u}}^{\mathrm{occ},\beta}
    S_{\bar{n}\bar{u}}
    \Big(
        \langle
            \mathbf{k}_i\bar{u}|p\bar{q}
        \rangle
        -
        \sum_r^{\mathrm{occ},\alpha}
            \langle 
                \mathbf{k}_i|r
            \rangle
            \langle 
                r\bar{u}|p\bar{q}
            \rangle
    \Big)
\end{alignat}

\section{Calculation of diabatic coupling in terms of basis functions}\label{App_dia_AO}
For the actual computation of the diabatic coupling, the MOs appearing in the electron-electron repulsion integrals in Eq. (\ref{vdia_mo_pw}) are further expanded with respect to the AO basis, giving rise to

\begin{widetext}
\begin{alignat}{1}
V^{\mathrm{dia}}_{i0}(\textbf{k}_i)
=&
\sum_n^{\mathrm{occ},\alpha} 
    \sum_{p}^{\mathrm{occ},\alpha}
        \sum_{q>p}^{\mathrm{occ},\alpha}
            (-1)^{s} \mathrm{det}\ \mathbf{S}_{in,pq} 
\sum_{\lambda \mu\nu}
    \left(
        c_\lambda^{(n)} 
        - 
        \sum_u^{\mathrm{occ},\alpha}
            c_\lambda^{(u)}S_{nu}
    \right)
    c_\mu^{(p)} c_\nu^{(q)} 
\Big[
    \langle 
        \mathbf{k}_i \lambda || \mu \nu 
    \rangle
    - 
    \sum_r^{\mathrm{occ},\alpha}
        \sum_{\rho\sigma} 
            c_\rho^{(r)} c_\sigma^{(r)} 
            \langle 
                \mathbf{k}_i | \rho 
            \rangle\ 
            \langle 
                \sigma \lambda || \mu \nu 
            \rangle
\Big]
\nonumber\\
+&
\sum_{\bar{n}}^{\mathrm{occ},\beta} 
    \sum_{p}^{\mathrm{occ},\alpha}
        \sum_{\bar{q}}^{\mathrm{occ},\beta}
            (-1)^{\bar{s}} \mathrm{det}\ \mathbf{S}_{i\bar{n},p\bar{q}} 
\sum_{\lambda \mu\nu}
    \left(
        c_\lambda^{(\bar{n})} 
        - 
        \sum_{\bar{u}}^{\mathrm{occ},\beta}
            c_\lambda^{(\bar{u})}S_{\bar{n} \bar{u}}
    \right)
    c_\mu^{(p)} c_\nu^{(\bar{q})} 
\Big[
    \langle 
        \mathbf{k}_i \lambda | \mu \nu 
    \rangle
    - 
    \sum_r^{\mathrm{occ},\alpha}
        \sum_{\rho\sigma} 
            c_\rho^{(r)} c_\sigma^{(r)} 
            \langle 
                \mathbf{k}_i | \rho 
            \rangle\ 
            \langle 
                \sigma \lambda | \mu \nu 
            \rangle
\Big]
\end{alignat}
where the Greek indices indicate Gaussian-type atomic basis functions.
\end{widetext}

Reordering the summations with respect to MO and AO indices, and defining
\begin{alignat}{1}
A_{\lambda\mu\nu}
=&
\sum_n^{\mathrm{occ},\alpha} 
    \sum_{q,p<q}^{\mathrm{occ},\alpha} 
        (-1)^{s} 
        \mathrm{det}\ 
        \mathbf{S}_{in,pq}
        \nonumber\\
        &
        \times\left(
            c_\lambda^{(n)} 
            - 
            \sum_u^{\mathrm{occ},\alpha}
                c_\lambda^{(u)} 
                S_{nu}
        \right)
        c_\mu^{(p)} c_\nu^{(q)}
\label{Afactor}
\\
\bar{A}_{\lambda\mu\nu}
=&
\sum_{\bar{n}}^{\mathrm{occ},\beta} 
    \sum_{p}^{\mathrm{occ},\alpha} 
        \sum_{\bar{q}}^{\mathrm{occ},\beta}
            (-1)^{\bar{s}} 
            \mathrm{det}\ 
            \mathbf{S}_{i\bar{n},p\bar{q}}
            \nonumber\\
            &
            \times\left(
                c_\lambda^{(\bar{n})} 
                - 
                \sum_{\bar{u}}^{\mathrm{occ},\beta}
                    c_\lambda^{(\bar{u})} 
                    S_{\bar{n} \bar{u}}
            \right)
            c_\mu^{(p)} c_\nu^{(\bar{q})}
\label{Abarfactor}
\\
B_{\sigma}
=&
\sum_r^{\mathrm{occ},\alpha}
    \sum_\rho
        c_\sigma^{(r)} 
        c_\rho^{(r)}
        \langle
            \textbf{k}_i | \rho
        \rangle
\label{Bfactor}
\end{alignat}
leads to the working equation

\begin{comment}

\begin{alignat}{1}
V^{\mathrm{dia}}_{i0}(\textbf{k}_i)
=
\sum_{\lambda \mu\nu}
    \Bigg(&
        A_{\lambda\mu\nu}
        \Big[
            \langle 
                \mathbf{k}_i \lambda || \mu \nu 
            \rangle
            \ -
            \sum_{\sigma} 
                B_{\sigma}
                \langle 
                    \sigma \lambda || \mu \nu 
                \rangle 
        \Big]
        +
        \nonumber\\
        &
        \bar{A}_{\lambda\mu\nu}
        \Big[
            \langle 
                \mathbf{k}_i \lambda | \mu \nu 
            \rangle
            \ -
            \sum_{\sigma} 
                B_{\sigma}
                \langle 
                    \sigma \lambda | \mu \nu 
                \rangle 
        \Big]
    \Bigg)
.\nonumber\\
\label{vdia_ao_main}
\end{alignat}
\textcolor{red}{OR}
\end{comment}
\begin{alignat}{1}
V^{\mathrm{dia}}_{i0}(\textbf{k}_i)
=
\sum_{\lambda \mu\nu}
    \Bigg(&
        \Big[
            \langle 
                \mathbf{k}_i \lambda | \mu \nu 
            \rangle
            \ -
            \sum_{\sigma} 
                B_{\sigma}
                \langle 
                    \sigma \lambda | \mu \nu 
                \rangle 
        \Big]
        \big(
            A_{\lambda\mu\nu}
            +
            \bar{A}_{\lambda\mu\nu}
        \big)
        \nonumber\\
        -&
        \Big[
            \langle 
                \mathbf{k}_i \lambda | \nu \mu 
            \rangle
            \ -
            \sum_{\sigma} 
                B_{\sigma}
                \langle 
                    \sigma \lambda | \nu \mu 
                \rangle 
        \Big]\ 
        A_{\lambda\mu\nu}
    \Bigg)
.\nonumber\\
\end{alignat}

\section{Spreading of a freely propagated LCAO-wavepacket}\label{App_C}
For the approximate description of adiabatic ionization processes as discussed in subsection \ref{el_ion}, we consider the HOMO $\phi(\mathbf{r})$ of anionic vinylidene as the initial free-electron wavepacket and compute its $\hat{\mathbf{r}}^2$ expectation value during free propagation,
\begin{equation}
    \langle \hat{\mathbf{r}}^2\rangle(t) = \langle \phi(\mathbf{r},t) |\hat{\mathbf{r}}^2 |\phi(\mathbf{r},t)\rangle = \sum_{\mu\nu} c_\mu c_\nu \langle\varphi_\mu(\mathbf{r},t)|\hat{\mathbf{r}}^2| \varphi_\nu (\mathbf{r},t)\rangle, \label{expect_r2}
\end{equation}
where $\varphi_{\mu,\nu}$ denote the time-propagated Gaussian atomic basis functions. Employing the free propagator $K(\mathbf{r},\mathbf{r}',t,0)=\langle \mathbf{r} | \mathrm{exp}(-i\hat{\mathbf{p}}^2 t/2m_e\hbar)|\mathbf{r}'\rangle $ these can be calculated as 
\begin{eqnarray}
\varphi_\mu(\mathbf{r},t) = \int d^3\mathbf{r}\, K(\mathbf{r},\mathbf{r}',t,0) \varphi_\mu (\mathbf{r},0).
\end{eqnarray}
For Cartesian basis functions of $s$, $p$ and $d$ type at the center $\mathbf{A}$, with the angular momentum quantum numbers $l$, $m$ and $n$ for the three spatial dimensions, the following analytic expressions are obtained:
\begin{eqnarray}
  \varphi_\mu(\mathbf{r},t) &=& N_{lmn} \mathrm{e}^{-\frac{3i\pi}{2}} (1+i\beta t)^{-(l+m+n+\frac{3}{2})} \mathrm{e}^{-\frac{\alpha}{1+i\beta t}\mathrm{r}^2} \nonumber \\
  &\times& (x-A_x)^l(y-A_y)^m(z-A_z)^n
\end{eqnarray}
if $l$, $m$, $n$ assume values of 0 or 1 and
\begin{eqnarray}
 \varphi_\mu(\mathbf{r},t) &=& N_{lmn} \mathrm{e}^{-\frac{3i\pi}{2}}  \mathrm{e}^{-\frac{\alpha}{1+i\beta t}\mathrm{r}^2} \left [-\frac{i\beta t}{2\alpha}(1+i\beta t)^{-\frac{5}{2}}\right. \nonumber \\
  &-& \left. (1+i\beta t)^{-\frac{7}{2}} (x-A_x)^l(y-A_y)^m(z-A_z)^n\right ]
\end{eqnarray}
if one of the numbers $l$, $m$, $n$ equals 2 and the others are zero. In the above expressions, $\alpha$ denotes the basis function exponent and $\beta=\frac{2\hbar\alpha}{m_e}$. The basic AO integrals occuring in Eq. (\ref{expect_r2}) can be calculated with common alorithms such as the McMurchie-Davidson scheme.\cite{mcmurchie78}
\nocite{*}
\bibliography{aid}% Produces the bibliography via BibTeX.

\end{document}